\documentclass[a4paper,12pt,oneside]{report}
\usepackage[MScThesis,lablogo]{EPFLreport}
\usepackage{xspace}
\usepackage{tocbibind}
\microtypecontext{spacing=nonfrench}
\providecommand{\tightlist}{%
  \setlength{\itemsep}{0pt}\setlength{\parskip}{0pt}}

\title{SVSHI: Secure and Verified Smart Home Infrastructure}
\author{Andrea Veneziano \& Samuel Chassot}
\adviser{Prof. George Candea}
\expert{Phillip Schanely}

\dedication{
\begin{raggedleft}
    A common mistake that people make when trying to design something completely foolproof is to underestimate the ingenuity of complete fools.\\~\\
    --- Douglas Adams\\
\end{raggedleft}
}

\acknowledgments{
We would like to thank Prof. Candea for his invaluable supervision and help. \\

\noindent We would also like to thank our families, who made all of this possible, and our girlfriends, whose support and patience have been essential to our studies. \\
}

\usepackage{graphicx}
\graphicspath{ {res} }

\usepackage[frozencache]{minted}

\usepackage{csquotes}

\makeatletter
\let\blx@rerun@biber\relax
\makeatother

\begin{document}
\maketitle
\makededication
\makeacks

\begin{abstract}

Smart infrastructures uses are growing and with them the need for dependability and correctness. To provide better correctness guarantees and bring formal verification into the equation, we present SVSHI, a platform for developing, verifying, and running Python applications in KNX installations, one of the most used smart buildings standards. SVSHI leverages abstract syntax tree (AST) manipulation, code generation, symbolic execution, and static configuration verification to make writing advanced apps easy, quick, and safe. With SVSHI, the reliability and compatibility of the applications are guaranteed without foregoing users' productivity.

\end{abstract}

\maketoc

\listoffigures

\chapter{Why does SVSHI exist?}

The source code is available
\href{https://github.com/dslab-epfl/svshi}{online}\footnote{\url{https://github.com/dslab-epfl/svshi}}. This report is based on the repository's white paper.

\section{Introduction}
\textbf{Smart infrastructures} are becoming more and more popular nowadays \cite{PrecedenceSmartBuildingMarket}. They allow the interconnection of all of the building's devices such as HVAC (Heating, ventilation, and air conditioning), blinds, windows, lights, PAC (Programmable Automation Controller), security, health monitoring, and switches to implement complex behaviors and automate tasks based on measurements and physical feedback. \\
Smart buildings can be commercial as well as domestic buildings, such as hospitals, factories, and houses. \\
Additionally, the Internet of Things advent is also contributing to this field's growth, with an ever-increasing amount of connected and affordable devices.

Several protocols and standards are used in smart buildings. They are divided into wireless and wired protocols. In the former set, we can find Wi-Fi, Bluetooth, Zigbee\footnote{\url{https://csa-iot.org}} and
Z-Wave\footnote{\url{https://www.z-wave.com}}, while in the latter we find mainly KNX\footnote{\url{https://www.knx.org/knx-en/for-professionals/What-is-KNX/A-brief-introduction/}},
Insteon\footnote{\url{https://www.insteon.com}} and
Velbus\footnote{\url{https://www.velbus.eu/domotica/}}.
KNX is one of the most widely used for large installations with thousands of devices, such as the industrial ones; wireless protocols are deployed instead primarily for more "toy" installations with few devices inside homes, as they can interact easily with smart home assistants such as Google Home and Amazon Alexa.

Becoming increasingly similar to a computer, the building has to face analogous challenges in terms of security, safety, reliability, and correctness. Even the most stable, mature, and used standard \cite{knx-dominant}, KNX, is lacking in this area, with severe issues (see Section \ref{current-issues}) that hinder its usability and dependability when programming advanced systems.

This project stems from the desire of solving these issues and demonstrating that reliable and complex applications can be easily written for smart buildings, with the end goal of offering a 100\% secure and verified platform to program smart infrastructures.

In this work, we thus present \textbf{SVSHI} (\textbf{S}ecure and \textbf{V}erified \textbf{S}mart \textbf{H}ome \textbf{I}nfrastructure) (pronounced like "sushi"), a platform/runtime/toolchain for developing and running formally verified smart infrastructures, such as smart buildings, smart cities, etc.

Our main contributions consist of an open-source project that empowers KNX professionals and hobbyists to develop and run Python applications that are formally verified in smart infrastructures, together with 3 real-life prototypes that exemplify what is currently achievable. To our knowledge, this is the first work to investigate the formal verification of software running in smart buildings and to propose a platform to program KNX installations in a high-level language.

\hypertarget{risk-stories}{%
\section{Risk stories}\label{risk-stories}}

In this section, we present two risk stories that highlight potential problems due to misconfiguration in smart infrastructures and the importance of their verification.

\hypertarget{hot-water-temperature-issues-in-homes}{%
\subsection{Hot water temperature issues in
homes}\label{hot-water-temperature-issues-in-homes}}

The water temperature inside domestic water heaters presents two conflicting risks: the risk of contracting the Legionnaires’ disease\footnote{\url{https://en.wikipedia.org/wiki/Legionnaires\%27_disease} (\textit{pulmonary legionellosis})}, caused by the proliferation of the \textit{Legionella} bacteria inside the tank, and the risk of scalding.

\textbf{Legionnaires' disease} is a dangerous illness, with death rates up to 12\% \cite{levesque2004}. The elderly, smokers, immunocompromised people, and patients suffering from chronic respiratory illnesses are the main groups at risk, and they include a large share of the population.

Residential drinking water supply is often the main cause for legionellosis \cite{alary1991} \cite{straus1996} \cite{stout1992}.

\textit{Legionella} proliferates in water at temperatures up to 45°C and does not replicate above 55°C. Therefore, the WHO recommends storing hot water at 60°C inside the water heater and to ensure, at least once a day, the temperature reaches at least 60°C in the entire tank \cite{who2002}.

Moreover, to avoid \textbf{scalding}, tap water should not exceed 49°C; this can be achieved with anti-scald devices \cite{levesque2004} that limit the temperature of the water at the faucet level.

Thus, a smart home needs to be properly configured to satisfy these requirements, as failure to do so could have severe consequences.

The importance of a system like SVSHI is therefore clear; \textbf{SVSHI can prevent the described problems} by:

\begin{itemize}
    \item Statically verifying that the programming of the KNX system respects the recommended temperatures.
    \item Automatically enforcing the safety condition in an app.
    \item Dynamically verifying (at runtime) that the recommended temperatures are preserved.
\end{itemize}

\hypertarget{co2-levels-in-buildings}{%
\subsection{CO2 levels in
buildings}\label{co2-levels-in-buildings}}

Another important issue is related to CO2 levels in buildings.

\textbf{CO2 levels} inside buildings such as offices and schools are very important and should be constantly monitored.

Increased indoor CO2 levels are strongly associated with symptoms such as fatigue, eye symptoms, headache, nasal symptoms, respiratory tract symptoms, and total symptom scores \cite{seppanen2000}. Also building ventilation rates are positively associated with the cited symptoms \cite{seppanen2000}. Furthermore, significant exposure-response relationships have been measured between CO2 concentration and respiratory symptoms, like sore throat, tight chest, irritated nose, and sinus combined mucous membrane symptoms and wheeze \cite{apte2000}.

Moreover, CO2 can be used as an indicator for poorly ventilated rooms, in which the concentration of microorganisms like viruses increases easily. This leads to higher chances of diseases spreading in a room\footnote{\url{https://tinyurl.com/gov-uk-covid}}, which has been put on the spot during the COVID-19 pandemic.

A possible solution to this problem is to keep indoor CO2 concentrations below 1000 ppm, the American Society of Heating, Refrigerating and Air-Conditioning Engineers (ASHRAE) action level\footnote{\url{https://www.ashrae.org/File\%20Library/Technical\%20Resources/Standards\%20and\%20Guidelines/Standards\%20Addenda/62.1-2016/62_1_2016_d_20180302.pdf}}.

\textbf{SVSHI can be easily used to this end}, by:

\begin{itemize}
    \item Statically verifying that the programming of the KNX system respects the threshold.
    \item Automatically enforcing the safety condition with an app.
    \item Dynamically verifying (at runtime) that the threshold is never reached.
\end{itemize}

\hypertarget{knx-background}{%
\section{KNX background}\label{knx-background}}

\textbf{KNX} is a communication protocol designed to interconnect buildings' devices. The basic idea is to connect all devices to a bus on which they can exchange packets of information called \textit{telegrams}. Devices can then send or receive telegrams and react to them. \\
Various physical links technologies can be used for the bus, the most used being a twisted pair cable. Other possible physical links are radio frequencies for wireless implementations, IP (so Ethernet cables or Wi-Fi), and power lines (i.e., used modulation of the electrical current in a standard electrical installation).

\hypertarget{knx-topology}{%
\subsection{KNX topology}\label{knx-topology}}

The KNX bus is split into different parts according to a particular \emph{tree} topology. That means that the bus cannot form a cycle.

The basic structure is the \textbf{line}. Each line consists of a power supply, a line repeater (if the complete bus contains more than one line), and up to 64 devices (including repeaters). It is theoretically possible to extend a line to contain up to 256 devices using more repeaters but it is better practice to split the bus into multiple lines if those many devices are needed.

Up to 15 lines can then be coupled together to form an \textbf{area}. An area line links together area couplers to which lines are connected.

Up to 15 areas can be coupled to the \textbf{main line} which forms the backbone of the installation. The main line and the areas form a complete system.

This architecture helps manage the system at a human level by compartmentalizing the devices into distinct groups. This also allows the repeaters and couplers to filter telegrams according to their destination addresses to reduce traffic.

An \textbf{individual address} (also called \textbf{physical address}) is assigned to each device on the bus following the topology above. This address is of the form: \textbf{A.L.D} where \textbf{A} is the number of the area, \textbf{L} is the number of the line in the area, and \textbf{D} is the number of the device on the line.
In binary, the individual address is therefore of the form: \textbf{AAAA.LLLL.DDDDDDDD} since A's and L's values are between 0 and 15, and D's between 0 and 255.
Couplers always have the number 0 on their respective line (\textbf{A.L.0}).

KNX over IP can be used in place of the main line and of area lines. Lines connected to devices must be twisted pairs (or radio frequencies for compatible devices). This is done using KNXnet/IP interfaces that have one Ethernet port and one connection to the bus.
KNX over IP can be used to interconnect multiple buildings together, e.g. for public administrations.
Most KNXnet/IP interfaces nowadays support tunneling mode which means that they can be used to program devices using ETS (see Section \ref{configuration-and-ets}).

\begin{figure}[H]
\begin{center}\includegraphics[scale=0.2]{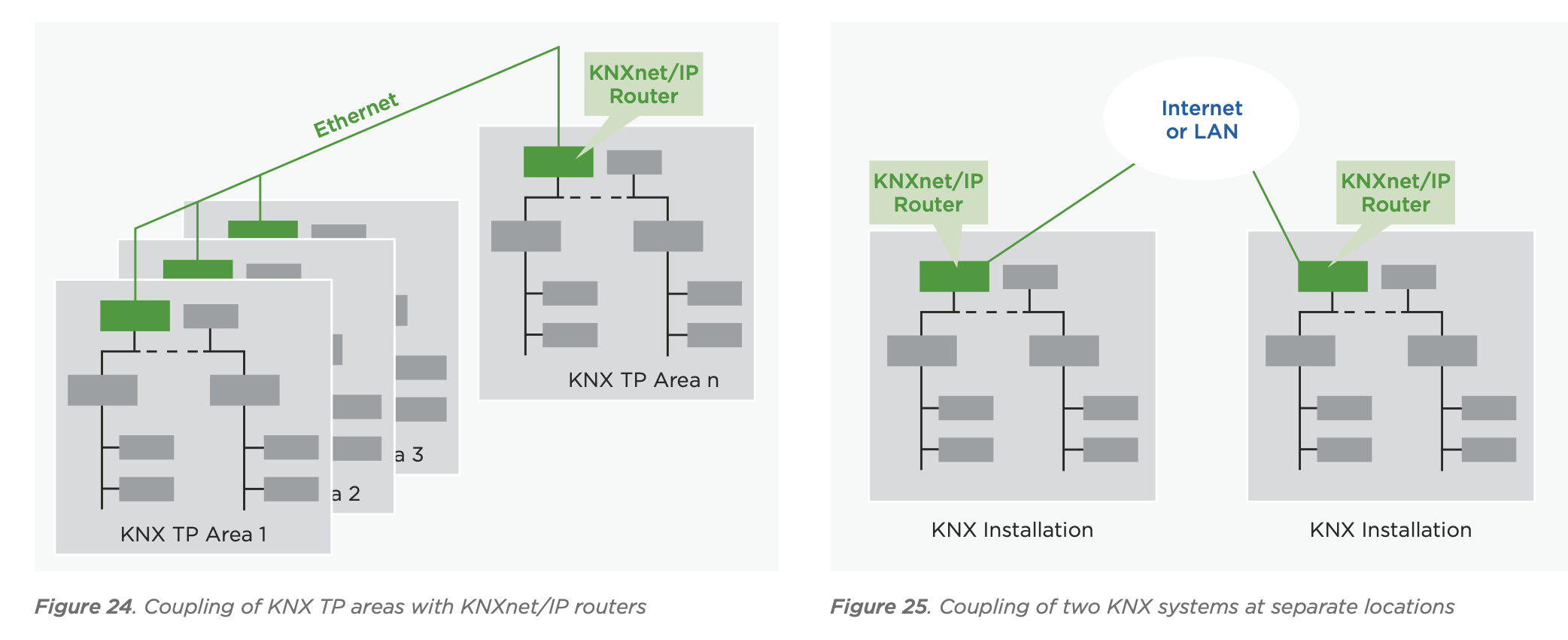}
\caption{Example of use of IP in KNXnet \cite{knx-basics}.}
\label{knx_topology_example}
\end{center}
\end{figure}

\begin{figure}[H]
\begin{center}\includegraphics[scale=0.2]{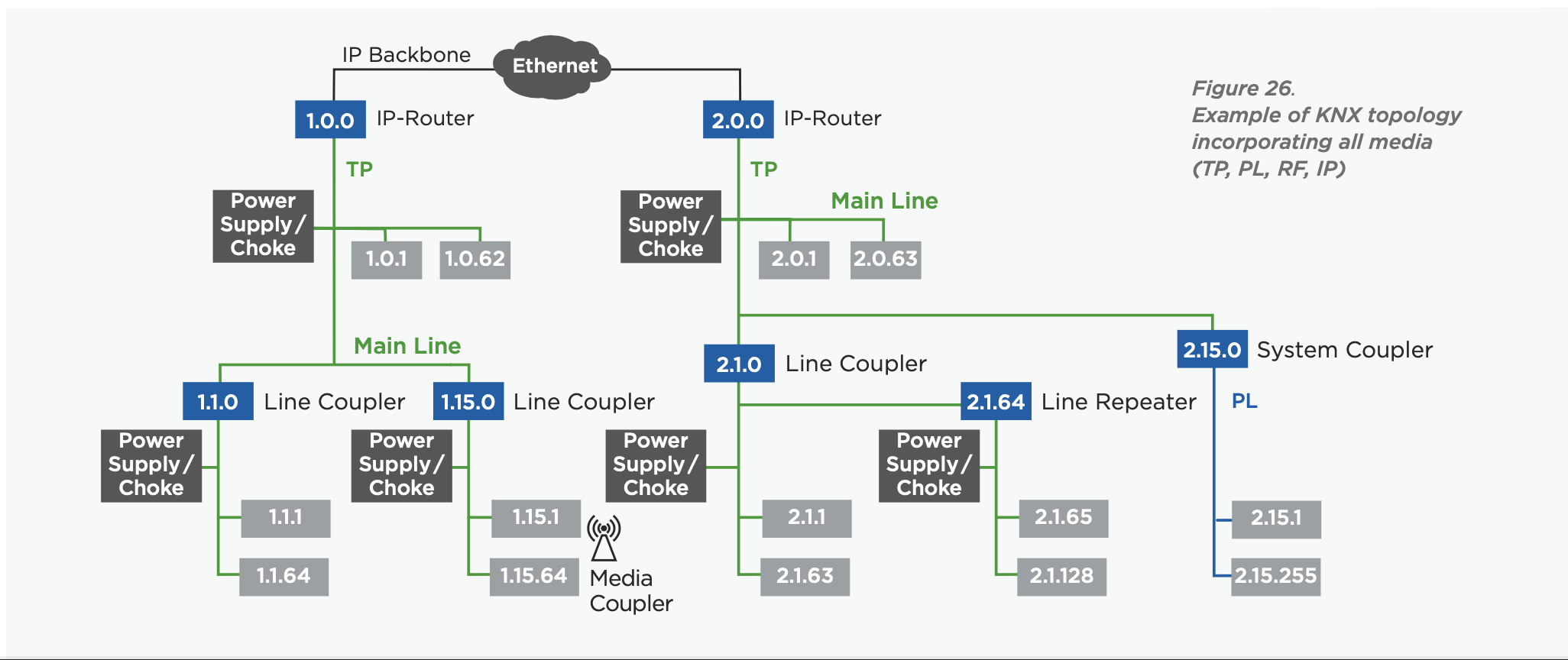}
\caption{Example of a complete KNX topology \cite{knx-basics}.}
\label{knx_complete_topology}
\end{center}
\end{figure}

Figure \ref{knx_complete_topology} shows an example of a KNX topology mixing up different technologies together.

\hypertarget{knx-telegram}{%
\subsection{KNX telegram}\label{knx-telegram}}

\textbf{KNX telegrams} are the information packets that transit on the KNX bus. Sensors send telegrams and actuators react to them.

\begin{figure}[H]
\begin{center}\includegraphics[scale=0.4]{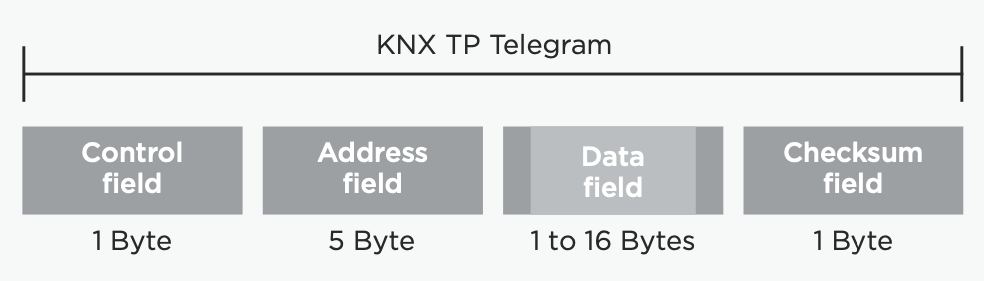}
\caption{Structure of a KNX telegram over twisted pairs \cite{knx-basics}.}
\label{telegram-fig}
\end{center}
\end{figure}

The telegrams' structure varies from one physical link to another. Here we develop only the structure of telegrams over twisted pairs.
A telegram is structured in fields, each field is composed of a certain number of bytes (as shown in Figure \ref{telegram-fig}):

\begin{itemize}
\tightlist
\item
  \textbf{Control field}: contains the priority of the telegram and whether or
  not it was repeated
\item
  \textbf{Address field}: specifies the \textbf{individual address} of the sender
  and the destination address (\textbf{individual address} or
  \textbf{group address}, see next section)
\item
  \textbf{Data field}: contains the telegram's payload (up to 16 bytes)
\item
  \textbf{Checksum field}: checksum for the parity check
\end{itemize}

\hypertarget{functionality-and-group-addresses}{%
\subsection{Functionality and group
addresses}\label{functionality-and-group-addresses}}

As previously said, devices can send and receive \textbf{KNX telegrams}
over the \textbf{KNX bus}. Now we develop how devices are programmed to interact together to implement a particular piece of functionality.

First of all, the KNX system works in a completely decentralized way: devices send and receive telegrams on their own without any centralized authority regulating the exchange. That means that a device failure does not impact devices that are not relying on the failing one to do their job.

To implement functionality, the KNX protocol offers the concept of 
\textbf{group addresses}. These are multicast addresses. This means that a device is configured to send a telegram to a particular group address at a specific event (e.g., when a push-button is pressed, send "on" or "off") and other devices that are programmed to react to this same group address will process the telegram.
Devices ignore telegrams that target group addresses they are not programmed to react to.

Group addresses can be seen as \textit{virtual cables} linking devices together to implement functions. Each group address corresponds to ONE function. To understand this, let us develop an example: in an office, two lights work independently. We want to be able to turn on or off light 1 and light 2 individually but also both together with another button. There are then 3 \textit{functions}: "light 1 on/off", "light 2 on/off“ and "light 1 and 2 on/off". Each of them must be assigned to a different group address. Then, buttons can be assigned to one of the three depending on what they should do. Lights are programmed to respond to 2 group addresses each.

Group addresses have multiple possible styles: 3-levels, 2-levels or
free. Each project uses only one style for its group addresses.
Regardless of the style, there are \textbf{65 536} possible group addresses
\cite{knx-basics}.

The most used style is the 3-levels. Using it, group addresses have the
following form: \textbf{X/Y/Z} where X represents the \emph{main group},
Y the \emph{middle group} and Z the \emph{sub group}. In this mode, the
ranges are \cite{knx-basics}:

\begin{itemize}
\tightlist
\item
  $0 \leq X \leq 31$,
\item
  $0 \leq Y \leq 7$,
\item
  $0 \leq Z \leq 255$
\end{itemize}

In the 2-level mode, there are only the \emph{main} and \emph{sub group}
with the following ranges \cite{knx-basics}:

\begin{itemize}
\tightlist
\item
  $0 \leq X \leq 31$,
\item
  $0 \leq Z \leq 2047$
\end{itemize}

Using the free style, the user can create an arbitrary number of
\emph{ranges} with the capacity she wants. These ranges can be nested. The
number of addresses still cannot exceed 65 536.

\newpage

\hypertarget{knx-devices}{%
\subsection{KNX devices}\label{knx-devices}}

The name ``\textbf{KNX device}'' represents any device that can be
connected to the bus. There are two types:

\begin{itemize}
\tightlist
\item
  \textbf{System devices}: these are the devices part of the KNX
  backbone, useful to the KNX system itself like power supplies,
  couplers, USB interfaces, KNXnet/IP interfaces, etc.
\item
  \textbf{End devices}: these are the devices that use KNX to work
  (sensors and actuators)
\end{itemize}

There exist two categories of end devices:

\begin{itemize}
\tightlist
\item
  \textbf{Sensors}: devices that detect events and transmit information on the
  bus
\item
  \textbf{Actuators}: devices that react to information on the bus and modify
  some physical world state
\end{itemize}

A majority of devices nowadays fall in both categories. For example, an HVAC (Heating Ventilation Air Conditioning) controller can modify the HVAC system state and also retrieve information about the system like current water temperature and send it on the bus. Thus it is a sensor and an actuator at the same time.

\begin{figure}[H]
\begin{center}\includegraphics[scale=0.4]{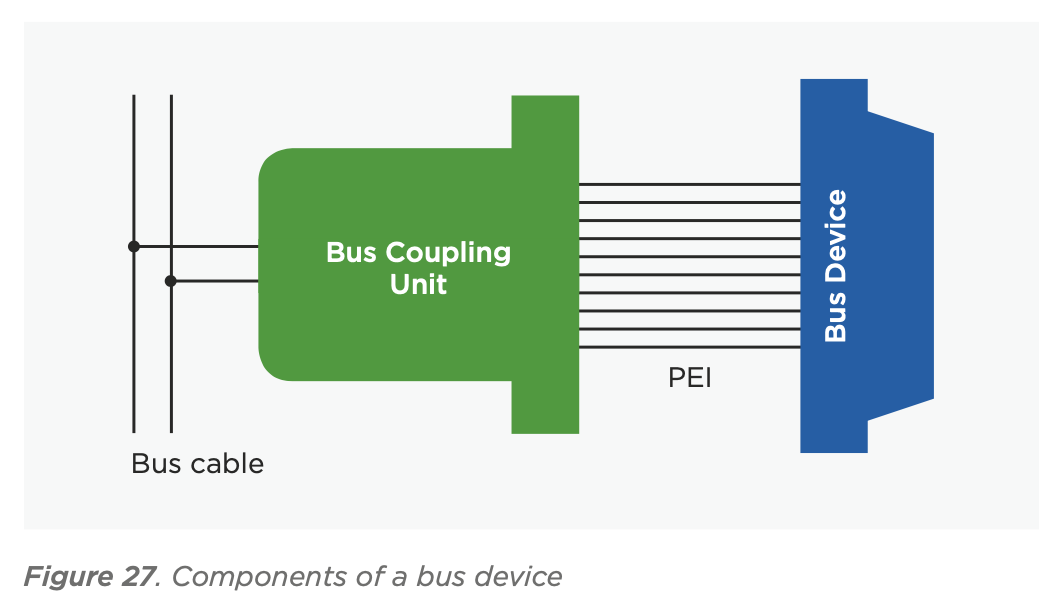}
\caption{Structure of an end device \cite{knx-basics}.}
\label{end-device-fig}
\end{center}
\end{figure}

An end device is composed of two parts: the \emph{Bus Coupling Unit} (or
BCU) and the \emph{Bus Device} (Figure \ref{end-device-fig}). The BCU is responsible for the
communication on the bus and thus of the KNX protocol implementation while
the Bus Device is the device itself (e.g., a push-button, a light
actuator, etc.). They are connected by a \emph{Physical External
Interface} (or PEI) of generally 10-12 pins.

\begin{figure}[H]
\begin{center}\includegraphics[scale=0.4]{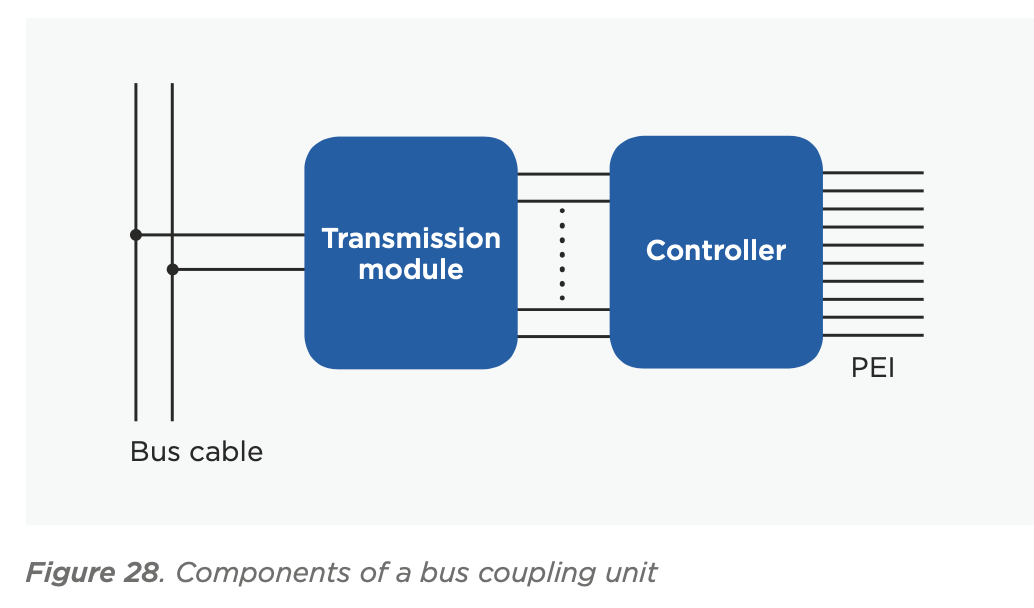}
\caption{Details of the Bus Coupling Unit \cite{knx-basics}.}
\label{bcu-fig}
\end{center}
\end{figure}

The BCU itself is composed of a \emph{Transmission module} and a
\emph{Controller} (Figure \ref{bcu-fig}). The Transmission module determines through which medium the device can be connected to the bus (i.e., it is different for each bus physical link technology); the controller is a microcontroller responsible for the implementation of the KNX protocol.

\newpage

End devices are manufactured by a large variety of manufacturers and can be simple or advanced. A device of the same family (e.g., push button) can be very different from one manufacturer to another. Some devices offer multiple settings and features while others of the same nature are very basic.
For example, a push-button can be just a simple on/off sensor but it can also provide additional options like a timer to delay the sending of the state change on the bus (e.g., to turn off basements' lights only when the user arrives at the top of the stairs when the button is at the bottom). Moreover, the structure of the settings in ETS (see next section) can be very different from one brand to another because there are no clear guidelines.

\hypertarget{knx-datatypes}{%
\subsection{KNX datatypes}\label{knx-datatypes}}

KNX defines some datatypes (also called \texttt{DPT}). Each communication object has a datatype associated. Datatypes are named as follows: \textbf{DPT-X-Y}. The \textbf{X} corresponds to the type of the data: the format and the encoding. The \textbf{Y} defines the range and the unit of the value. For example, DPT-1 corresponds to boolean values, encoded on 1 bit. Then, the \textbf{Y} value in DPT-1-Y indicates whether the value is a \textit{switching} value or a \textit{move up/down} value for example.

Here is the list of the most used DPTs:

\begin{itemize}
\tightlist
\item
  1.yyy = boolean, like switching, move up/down, step
\item
  2.yyy = 2 x boolean, e.g.~switching + priority control
\item
  3.yyy = boolean + 3-bit unsigned value, e.g.~dimming up/down
\item
  4.yyy = character (8-bit)
\item
  5.yyy = 8-bit unsigned value, like dim value (0..100\%), blinds
  position (0..100\%)
\item
  6.yyy = 8-bit 2's complement, e.g.~\%
\item
  7.yyy = 2 x 8-bit unsigned value, i.e.~pulse counter
\item
  8.yyy = 2 x 8-bit 2's complement, e.g.~\%
\item
  9.yyy = 16-bit float, e.g.~temperature
\item
  10.yyy = time
\item
  11.yyy = date
\item
  12.yyy = 4 x 8-bit unsigned value, i.e.~pulse counter
\item
  13.yyy = 4 x 8-bit 2's complement, i.e.~pulse counter
\item
  14.yyy = 32-bit float, e.g.~temperature
\item
  15.yyy = access control
\item
  16.yyy = string -\textgreater{} 14 characters (14 x 8-bit)
\item
  17.yyy = scene number
\item
  18.yyy = scene control
\item
  19.yyy = time + data
\item
  20.yyy = 8-bit enumeration, e.g.~HVAC mode ('auto', 'comfort',
  'standby', 'economy', 'protection')
\end{itemize}

\hypertarget{knx-simulator}{%
\subsection{KNX simulator}\label{knx-simulator}}

The KNX Association offers a \textit{simulator} for a KNX system named \textit{KNX Virtual} \cite{knx-virtual}. It is a Windows-based application that emulates a KNX installation. It can be used to learn the KNX terminology and get familiar with the technology.
Furthermore, it can be leveraged to get used to ETS as well: ETS can connect to it and program the devices. Then, a GUI lets the user play with the sensors and shows the result on the simulated actuators.

The available devices are pre-installed and cannot be changed. The software GUI offers several views with different devices shown.

The devices are available on ETS through a catalog similar to the ones offered by the "real" manufacturers.

The main downside of this simulator is the fact that the devices are unique to KNX Virtual. The devices available do not mirror real devices and thus the configuration is simplified compared to a real device's one. Moreover, this cannot be used to simulate an existing installation, for example, to test the solution we developed.

KNX Virtual can be used to play with SVSHI though. A developer can test her application developed for SVSHI if the devices it uses are available on KNX Virtual. This does not fit all cases but it is useful to familiarize with the platform and for the SVSHI developers who just want to test manually the platform itself.

Other commercial solutions exist like \textit{KNX Simulator}\footnote{\url{https://www.knxsimulator.com}} but as they are not free, we did not explore them in detail.

\hypertarget{configuration-and-ets}{%
\subsection{Configuration and ETS}\label{configuration-and-ets}}

To configure KNX devices, the person installing a KNX system (we call her the "programmer") needs to use ETS. \textbf{ETS} is a piece of software provided by the KNX Association. It is closed source and not free.

Each manufacturer provides a KNX catalog entry for each of its devices.

The programmer then represents the whole system in ETS, following the topology formed by the devices in the real building. Each device added in ETS must correspond to the exact model installed in the building (using the manufacturer's catalog entry).

Once each device is entered in ETS, the programmer configures them to achieve the wanted behaviour for the building. She creates a group address for each function (as explained in Section \ref{functionality-and-group-addresses}).

Each device exposes \textit{communication objects} that are like IO ports for the device. Each communication object can be linked to one or more group addresses.
For example, a push-button will at least expose a communication object "on/off state" that is like an output port connected to the bus that writes to the configured group address every time the state changes. The communication objects can be input, output, or both.

Devices also have settings that have to be modified in ETS. These settings can include parameters about the frequency at which updates are sent to the bus, whether or not the feedback should be sent (i.e., for an actuator sending its current state to the bus), complex settings about its behaviour, etc. These settings are, as explained by the domotic expert we met, the trickiest part about the configuration of a KNX installation. They are designed freely by the manufacturers, meaning that there is no uniformity across brands. Programmers must then learn how to configure each device by reading the documentation (also no uniformity there), which takes a lot of time.

One important aspect to note is that the rules that dictate the behaviour are never represented or stored in ETS. The rules are generally written down in some documents the programmer receives from the client (or designs herself) but they do not have any machine-readable representation related to ETS. This means that only having the ETS project and the configured devices, it is not possible to get back the rules. There is also no direct relation between the rules and the actual configuration in ETS, therefore errors can easily be made by doing the "translation", which can lead to unwanted behaviour of the final installation.

Once the configuration in ETS is done, the programmer can download to each device its settings. The first time a device is configured, it must receive its individual address (see Section \ref{knx-topology}). For this step, a button must be physically pressed on the device to allow the individual address setting. For consequent application downloads, this is not necessary. ETS downloads the configuration to each device following the manufacturer's catalog entry by writing values in the device memory through commands sent on the bus.

After the application download is done, devices start sending and reacting to telegrams and the building is configured.

One should never lose the ETS project file because it is impossible to infer it from the devices. Thus, if the project file is lost, everything has to be done again from scratch.

\hypertarget{knx-pros-and-cons}{%
\subsection{KNX Pros and Cons}\label{knx-pros-and-cons}}

In this section, we develop the advantages and disadvantages of KNX as a smart buildings' backbone.

The biggest advantage of KNX is its reliability. The protocol indeed relies on simple electrical cables, which are simpler than Ethernet ones, for instance. The installation is almost as stable and reliable as a traditional electrical installation. Another aspect that improves reliability is the certification system: each device is tested and approved by the KNX Association to get the "KNX-certified" logo \cite{knx-certified}. KNX-certified devices are thus tested by a third party and not just the manufacturer, which helps improve the quality and reliability of devices.

Another advantage of KNX is its distributed architecture. This means that if a device fails, the rest of the system that does not depend on this device continues to work normally.

The KNX protocol has some downsides too.

The first major one is the price. Devices cost generally more than similar devices for traditional installations. This means that the cost of a KNX installation is greater than the one of a standard electrical installation, which leads to a low adoption rate for domestic usage.

Even if the KNX bus over radio frequencies exists in the specification, in practice, there exist almost no compatible devices. The KNX system uses thus mainly cables. This can be a problem in some special buildings but it is mainly an obstacle to retrofitting. It is indeed painful to install the bus cable in a building where it was not done at construction time.

Configuration is also an issue: it must be done using ETS, which is time-consuming and cumbersome to use. Moreover, the translation of the rules to devices settings is completely non-trivial. Experience with the system and the devices is required to be able to exploit their full potential and to implement the desired functionality. Therefore, an expert is often required to configure even simple buildings like houses and thus, end users are not confident enough to do modifications to their own installations.

The KNX certifications are a good point for reliability but they lead to business exploitations that are a disadvantage for the hobbyists. The licenses and certifications are expensive and the association has the habit of giving them only to professional electricians \cite{knx-alternatives-forum}. Therefore, the public lacks documentation and needs to buy proprietary software to use an open protocol and system.

\hypertarget{why-knx}{%
\subsection{Why KNX?}\label{why-knx}}

In this section, we develop why we decided to develop a solution around KNX.

Three main points of consideration make us argue that KNX is the protocol to use for such a project.

Firstly, KNX is the most widely used system for smart buildings \cite{PrecedenceSmartBuildingMarket} \cite{knx-market-share} and one of the oldest. The KNX standard has indeed existed since 1999 and inherits from the EIB (European Installation Bus) system which was created in 1987 \cite{knx-basics} \cite{bus-eib-wikipedia}. Nowadays, KNX represents, for example, 56\% of the market of smart buildings protocols in Germany, 42\% in China, 34\% in the Netherlands, and 27\% in the UK \cite{knx-market-share}.

Secondly, as KNX is mainly a wired protocol built directly in the buildings' walls, we argue that the installations are meant to be used for many decades. The reliability and simplicity of the protocol also play a major role in the perenniality of the protocol.

Finally, as we are aiming for reliable and verified smart buildings infrastructures, we need a reliable protocol on which to rely. As explained in Section \ref{knx-pros-and-cons}, the KNX system is simple and reliable and thus represents a sane base for our new system.

Given the adoption and features of the system, we argue that KNX is a central building block for the future of smart buildings. Moreover, its perenniality and reliability make KNX a better choice than consumer-grade technologies like Google Home\footnote{\url{https://developers.google.com/assistant/smarthome/overview}} or Apple HomeKit\footnote{\url{https://developer.apple.com/homekit/}} which might not exist anymore in their current form (if at all) a few years from now. This is even more important in the case of public administration or facilities buildings that our system is also targeting. For these types of buildings, consumer-grade systems are indeed not an option.

On another front, KNX is not a popular research subject. We could not find significant current research work and we, therefore, see an opportunity to propose something new and valuable for this technology.

\hypertarget{current-issues}{%
\section{Current issues}\label{current-issues}}

KNX is stable and well established. We however identified three main issues which slow down the adoption rate and what is achievable using KNX. We expose and explain here these three points.

\hypertarget{time-consuming-configuration}{%
\subsection{Time-consuming configuration}\label{time-consuming-configuration}}

The configuration of a KNX installation is time-consuming. As we explained in Section \ref{configuration-and-ets}, all the configuration takes place in ETS. This software is well developed but the way the configuration is performed is outdated: every device must be configured individually and nothing can be automated. Moreover, as each device's settings' structure is designed by the manufacturer, there exists no uniformity so the learning curve is steep.

As also explained earlier (Section \ref{configuration-and-ets}), the rules that dictate the devices' behaviour are not entered directly in the configuration. The translation between human described behaviours and the settings of the actual devices is done by hand. The two representations are completely different and thus the translation takes time and energy from the programmer.

The ETS project reflects \textit{exactly} the building's structure with all the devices and their locations, and thus is unique to the installation. That means that the work has to be done again from scratch (except the learning and experience accumulated, of course) every time the programmer begins a new project. It is a shame to not be able to reuse common configurations to gain time.

The configuration being that dissociated from the rules humans described, the probability of incorrect behaviours is high and so is the debug time. It is not rare that the programmer must come back to a project several times until everything settles.

\hypertarget{cumbersome-and-error-prone-project-evolution}{%
\subsection{Cumbersome and error-prone project evolution}\label{cumbersome-and-error-prone-project-evolution}}

The evolution of a project is important: a building can be extended or devices added, like a new HVAC system. Making a project evolvable is thus important and we think KNX does not provide a good enough solution.

First of all, the ETS project file must be accessible (Section \ref{configuration-and-ets}). If the file is lost, the configuration cannot be extracted from the devices (except the group addresses but these are only a fraction of the total configuration). A programmer must then discuss with the users and observe the behaviours of the current installation before trying to reconstruct the ETS project as a new one. This alone restricts evolution for existing buildings. We discussed with a professional KNX programmer that explained that for a house this process is doable but, for an industrial building, it is an impossible mission and he always refuses this kind of project.

Even if the ETS project file is available, evolution is not trivial. The new device must be added to the project and the new behaviour implemented. If the device works independently from the rest of the installation, this process is relatively simple, but if other existing devices are involved, it can become tricky. More specifically, if the new device is tightly bound to existing ones, the work is almost as challenging as for a new configuration.

Old behaviours must be, of course, preserved and this is not trivial. The issue is amplified by the fact that there is no way to detect that an existing behaviour has been wrongly modified by a new one; it must be tested in the real installation. This can become dangerous when the existing behaviour involves critical systems such as hot water heating (as described in Section \ref{risk-stories}), the ventilation of specific areas (e.g., hospital rooms), or access control for example.

\hypertarget{lack-of-tools-for-complex-behaviours}{%
\subsection{Lack of tools for complex behaviours}\label{lack-of-tools-for-complex-behaviours}}

Using ETS and KNX devices, the range of possible functionalities is limited. One can only connect devices' communication objects using group addresses and configure devices settings (e.g., some delays, sensors' refresh frequency, repeating data sent, etc.).

This is sufficient for a large number of situations like blinds automation with a weather station, lights management, etc. As long as a device exists to implement the intended behaviour, this method would work.

However, what about Machine Learning based behaviours or functions that would connect to external services? For now, this is not possible using existing solutions. There exist servers (like \textit{ComfortClick} \cite{confortclick}) that connect to the KNX bus and let the programmer develop more complex rules but the programming language is not a standard one and ML and external services are not available. These solutions are not numerous and some of the existing ones are in an end-of-life state (like \textit{Lifedomus} from DeltaDore \cite{lifedomus}).

There also exist logical modules but they allow the programmer to design logic circuits using gates on a GUI inside ETS to connect in/out pins of the device to group addresses. This is cumbersome and extremely low level so these are not a viable option for advanced behaviours.

We think that the KNX ecosystem lacks a way to program in established programming languages or at least a way to enter complex sets of rules directly. It also lacks a way of verifying that a configuration is safe and implements the desired behaviour and does not provide a tool to check that this behaviour is preserved through updates of the installation.

\hypertarget{how-do-we-solve-them}{%
\section{How do we solve them}\label{how-do-we-solve-them}}

We solve the issue related to time-consuming configuration mainly by \textit{reducing} the amount of time users need to spend on ETS. Most of the configuration is implicitly performed by developing the application, and SVSHI even outputs two files that describe what remains to be done on ETS after installing an application. Additionally, the rules describing the devices' behaviors are explicitly encoded in the apps by the user and automatically used and enforced while executing. Moreover, apps can be reused and adapted with minimal changes, without the need of re-doing all the work from scratch, like in ETS. Overall, since the probability of committing errors is decreased and the amount of code written is minimal, debugging and fixing mistakes become quick and simple tasks.

The second pain point, cumbersome and error-prone project evolution, is solved in multiple ways. First of all, a SVSHI distribution (including all installed apps) can be \textit{version-controlled}, which means that it will be always accessible and rollbacks in case of errors are trivial. Furthermore, changing (or adding) a behavior is as simple as modifying (or creating) a short Python script. Finally, every app update or new installation is verified to make sure it does not break the current system, so compatibility and correctness are always ensured.

The mere SVSHI existence overcomes the last problem, the lack of tools for complex behaviors. With SVSHI and limited programming skills, it is possible to use \textit{any} Python library available and write \textit{any kind} of apps needed; Machine Learning, external services, APIs are just some of the tools that can be leveraged in a SVSHI app with little effort and a high degree of abstraction.

\chapter{Related work}

This section contains a brief overview of the current smart
infrastructures state-of-the-art and SVSHI related work.

\hypertarget{commercial-and-open-source-software}{%
\section{Commercial and open-source software}\label{commercial-and-open-source-software}}

Few modern programming languages support communication with KNX using external libraries: Java (and more generally the JVM
ecosystem) with Calimero\footnote{\url{https://calimero-project.github.io/}},
Python with XKNX\footnote{\url{https://xknx.io/}} (the library SVSHI uses) and
Go with knx-go\footnote{\url{https://github.com/vapourismo/knx-go}}. They
allow to interact with KNX mainly via KNXnet/IP and offer different
degrees of abstraction.

Either on top of the above libraries or via custom-made solutions,
various commercial and open-source building automation software systems
have been built, such as Thinknx\footnote{\url{https://www.thinknx.com/v4/en/}},
ComfortClick\footnote{\url{https://www.comfortclick.com/}}, Weble\footnote{\url{https://www.weble.ch/}},
openHAB\footnote{\url{https://www.openhab.org/}} and Homey\footnote{\url{https://homey.app/en-us/}}.
They usually provide a central unit that integrates with different kinds of devices, systems, and standards, allowing to define rules and scenes. A typical use case is to create behavior that depends on both KNX and non-KNX devices. Compared to SVSHI, they need the KNX system to be already configured, they do not provide any form of formal verification, and complex behaviors and app invariants cannot be implemented in a high-level programming language such as Python.

SmartHomeNG\footnote{\url{https://www.smarthomeng.de}} is a metagateway that
interconnects several things together. It supports KNX and is used as a
gateway between several protocols. Other supported protocols include
HomeMatic, EnOcean, 1-wire, DMX, Philips Hue, MQTT, etc. It is written
in Python.

\hypertarget{academic-work}{%
\section{Academic work}\label{academic-work}}

Fu et Al. (2021) \cite{fu2021} work on \textit{dSpace}, a programming framework for smart space applications, shares some similarities with SVSHI in being a platform that simplifies application development for smart infrastructures and that decouples software and hardware. Although, compared to SVSHI, \textit{dSpace} does not provide formal verification, does not apply to KNX, and automation policies are not written in Python.

Research about KNX is rare and has been in decline in recent years. Even
rarer is research involving KNX and formal verification; an exception is
the paper by Shehata and Al. (2007) \cite{shehata2007}, who propose
a way to deal with policies that can interact badly and induce
unexpected behavior. Their solution is a run-time policy interaction
management module for detecting and resolving interactions among user
policies in KNX-based smart homes. It runs in the ETS software and is
used at programming time.

Another interesting work on the KNX standard was done by Ruta et Al. (2011) \cite{ruta2011}, who propose a modification
of the KNX stack to add a micro layer for semantic and enable autonomous
decisions by the devices rather than explicit user commands.

Publications concerning formal verification in smart buildings are more
common, compared to the KNX-related ones. Sun et Al. (2017) \cite{sun2017} present a lightweight
rule verification and resolution framework for verifying the correctness
of rules used by wireless sensor-actuator networks. It uses knowledge
bases and anomaly detection.

Trimananda et Al. (2020) \cite{trimananda2020} study conflicts between
apps on Samsung SmartThings\footnote{\url{https://www.smartthings.com/}}, a
platform for developing and deploying smart home IoT devices. Their
findings suggest that the problem of conflicts between smart home apps
is serious and can create potential safety risks. They also developed a
conflict detection tool that uses model checking to automatically detect
conflicts.

Grobelna et Al. (2017) \cite{grobelna2017} focus on the design and
verification methods of distributed logic controllers supervising
real-life processes. In their work, Petri Nets\footnote{\url{https://en.wikipedia.org/wiki/Petri\_net}}
are used to model the control system with all the transitions. The
resulting net is then formally verified with the application of model
checking techniques against predefined behavioral requirements.

Garcia-Constantino et Al. (2018) \cite{garcia-constantino2018} leverage Petri
nets too. They use them to model elderly daily activities such as
cooking pasta, making coffee and tea etc. Then, different scenarios are
verified to understand whether they could possibly indicate an abnormal
behavior, which can be used by the system to send alarms.

Corno et Al. (2011) \cite{corno2013} propose a design time
methodology for the formal verification of Intelligent Domotic
Environments. Their approach is based on modeling the system and its
algorithms with UML 2 State Charts\footnote{\url{https://sparxsystems.com/resources/tutorials/uml2/state-diagram.html}}
and formally verifying them by checking logical properties expressed in
temporal logic by model checking tools.

Machine learning applications in smart infrastructures are an important
research subject as well. Chatrati et Al. (2020) \cite{chatrati2020} propose a smart home health monitoring system that helps to analyze the patient’s blood pressure and glucose readings at home and notifies the healthcare provider in case of any abnormality detected. Hypertension and diabetes are predicted with conditional decision-making and machine learning approaches.

Li et Al. (2019) \cite{li2018} work, in the field of building
FDD (data-driven fault detection and diagnosis), consists of an expert
knowledge-based unseen fault identification method to identify unseen
faults by employing the similarities between known faults and unknown
faults.

Dey et Al. (2020) \cite{dey2020} present an extraction method
to deal with HVAC terminal unit. Machine learning is employed for fault
detection and diagnosis in building.

\chapter{Who is SVSHI for?}

This section details SVSHI's intended audience and how the public can
use SVSHI. It presents SVSHI from a user perspective. For more details
about how SVSHI works, please see
\protect\hyperlink{4-How-does-SVSHI-work}{Section 4}.

\hypertarget{whom-does-svshi-help}{%
\section{Whom does SVSHI help?}\label{whom-does-svshi-help}}

SVSHI helps both smart home enthusiasts and hobbyists wishing to tinker with their installation, and KNX professionals working on large business systems.

\hypertarget{how}{%
\section{How?}\label{how}}

\begin{figure}[ht]
\begin{center}\includegraphics[scale=0.4]{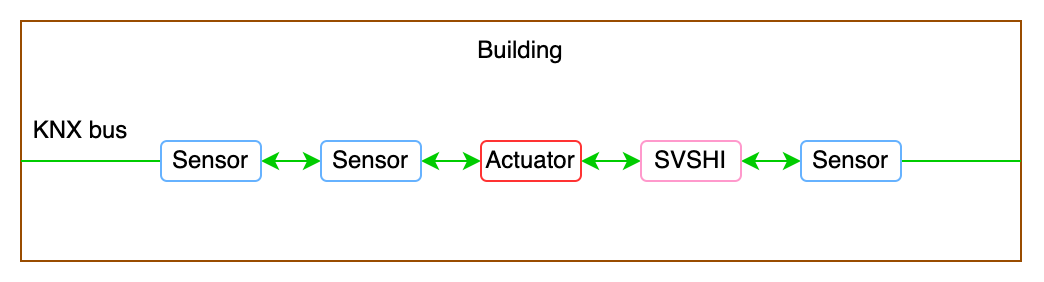}
\caption{SVSHI's place in a KNX installation.}
\label{svshi-in-knx-fig}
\end{center}
\end{figure}

As depicted in Figure \ref{svshi-in-knx-fig}, SVSHI acts like a KNX device, sending and reading telegrams from the bus.

Then, devices are configured to send all telegrams to SVSHI only and consequently, SVSHI sends telegrams to the devices to implement the functionality.

\begin{figure}[H]
\begin{center}\includegraphics[scale=0.25]{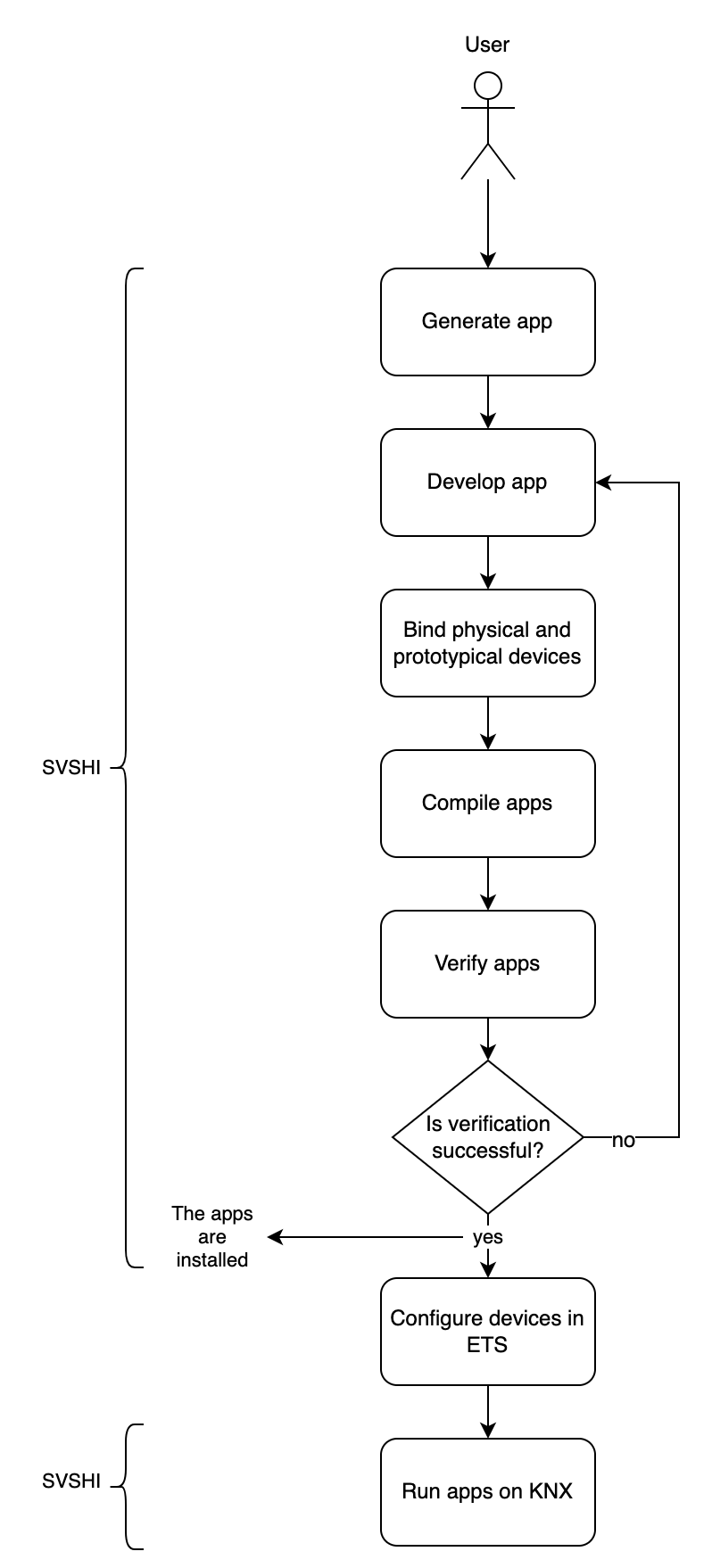}
\caption{Typical SVSHI workflow.}
\label{svshi-workflow-fig}
\end{center}
\end{figure}

In this way, SVSHI allows to develop and run Python applications in a KNX system. On top of that, it verifies that the apps adhere to a set of invariants provided by the user. No formal verification background is needed, the only requirement is some Python programming knowledge. Figure \ref{svshi-workflow-fig} presents a typical SVSHI workflow.

\newpage

Thus, with SVSHI smart home \textit{aficionados} can control their smart home KNX installation without having to rely necessarily on experts, and can add easily pieces of functionality by writing or downloading apps without the fear of breaking the system.

In a similar fashion, KNX professionals and domotics experts can tackle bigger, richer, and more complex projects, as the time gained with SVSHI is significant compared to an ETS-centric approach. Furthermore, the verification of the programs before deployment means no more surprises and no more afternoons of correcting errors. Finally, being able to add at any time pieces of functionality to the system by writing or downloading apps allows to smoothly evolve and improve projects, without headaches.

\hypertarget{developing-apps}{%
\subsection{Developing apps}\label{developing-apps}}

As explained in the previous section, developing apps with SVSHI is easy
and any Python library can be used. Additionally, the user has to spend
less time on ETS.

To develop an app for SVSHI, a user has to:

\begin{enumerate}
\def\labelenumi{\arabic{enumi}.}
\tightlist
\item
  Create the prototypical structure file containing the list of
  the devices the app should use, as explained in the
  \protect\hyperlink{prototypical-structure}{app prototypical
  structure} section (Section \ref{prototypical-structure}).
\item
  Run the app generator, as explained in the
  \protect\hyperlink{app-generation}{app generation} section (Section \ref{app-generation}), to get
  the app skeleton. It will be created under the \texttt{generated/}
  folder.
\item
  \protect\hyperlink{writing-apps}{Write the app}.
\item
  Run \texttt{svshi} to generate the bindings with
  \texttt{svshi\ generateBindings\ -f\ ets.knxproj}, where the argument
  is the \emph{absolute} path to the ETS project file.
\item
  Map the right physical ids given in
  \texttt{generated/physical\_structure.json} to the right device in
  \texttt{generated/apps\_bindings.json}. This is needed to provide the devices in the Python code with the group addresses to use. The first file represents the physical structure from the ETS project file, where each communication object has an id. The second one represents the apps' structures with the devices and for each of them, the links they need.
\item
  Run \texttt{svshi} again to
  \protect\hyperlink{compilation}{compile} and
  \protect\hyperlink{verification}{verify} the app with
  \texttt{svshi\ compile\ -f\ ets.knxproj}.
\end{enumerate}

For more details on how SVSHI works under the hood, see
\protect\hyperlink{how-does-svshi-work}{Section \ref{how-does-svshi-work}}.

\hypertarget{writing-apps}{%
\subsubsection{Writing apps}\label{writing-apps}}

To write an app, the user mainly has to modify the \texttt{main.py}
file, optionally adding dependencies into the \texttt{requirements.txt}
file provided in the generated project. To understand how the project is
generated, please refer to the
\protect\hyperlink{app-generation}{app generation section (Section \ref{app-generation})}.

All the available device instances are already imported in
\texttt{main.py}. They mirror what has been defined in the device
prototypical structure file.

The application can use external files. They however need to have been
declared in the prototypical structure json file and they have to be
located at the root of the project, next to \texttt{main.py}.

There are two important functions in \texttt{main.py},
\texttt{invariant()} and \texttt{iteration()}. In the first one, the user
should define all the conditions (or \emph{invariants}) that the entire
KNX system must satisfy throughout execution of \textbf{all}
applications, while in the second she should write the app code.

An important thing to be aware of is that \texttt{iteration()} cannot
use external libraries directly. Instead, these calls have to be defined
first inside \emph{unchecked functions}, which are functions whose name
starts with \texttt{unchecked} and whose return type is explicitly
stated. Then, these functions can be used in \texttt{iteration()}.

In addition, note that \texttt{invariant()} must return a boolean value,
so any kind of boolean expression containing the \emph{read} properties
of the devices and constants is fine. However, here operations with side
effects, external libraries' calls, and unchecked functions calls are
\textbf{not} allowed.

\textbf{Unchecked functions} are used as a compromise between usability
and formal verification, and as such must be used as little as possible:
their content is not verified by SVSHI. Furthermore, they should be
short and simple: we encourage developers to add one different unchecked
function for each call to an external library. All logic that does not
involve calls to the library should be done in \texttt{iteration()} to
maximize code that is indeed formally verified. \\
Nonetheless, the user can help the verification deal with their presence by annotating their docstring with \emph{post-conditions}.

Functions' \textbf{post-conditions} define a set of \emph{axioms} on the
return value of the function: these conditions are assumed to be always
true by SVSHI during verification. They are defined like this:
\texttt{post:\ \_\_return\_\_\ \textgreater{}\ 0}. You can use constants
and other operations. You can add as many post-conditions as you like
and need. Therefore, we encourage developers to avoid having
conjunctions in post-conditions but rather to have multiple
post-conditions. This does not make a difference for the verification but
helps the readability.\\
However, keep in mind that these conditions are \textbf{assumed to be
true} during formal verification! If these do not necessarily hold with
respect to the external call, bad results can occur at runtime even
though the code verification was successful!

An example with multiple post-conditions could be:

\begin{minted}{python}
def unchecked_function() -> int:
  """
  post: __return__ > 0
  post: __return__ != 3
  """
  return external_library_get_int()
\end{minted}

Furthermore, applications have access to a set of variables (the
\emph{app state}) they can use to keep track of state between calls.
Indeed, the \texttt{iteration()} function is called in an
\protect\hyperlink{execution}{event-based manner} (either the KNX
devices' state changes or a periodic app's timer is finished, see Section \ref{execution}). All calls
to \texttt{iteration()} are independent and thus SVSHI offers a way to
store a state that will live in between calls. There is a local state
instance \emph{per app}.

To do so, in \texttt{main.py} the \texttt{app\_state} instance is
imported along with the devices. This is a \textit{dataclass}\footnote{\url{https://docs.python.org/3/library/dataclasses.html}}
and it contains 4 fields of each of the following types: \texttt{int},
\texttt{float}, \texttt{bool} and \texttt{str}. The fields are called
respectively \texttt{INT\_X}, \texttt{FLOAT\_X}, \texttt{BOOL\_X} and
\texttt{STR\_X} where X equals 0, 1, 2, or 3. We made this design choice
to simplify the verification, as explained in
\protect\hyperlink{crosshair-in-svshi-and-code-modification}{Section
\ref{crosshair-in-svshi-and-code-modification}}.

These values can be used in \texttt{iteration()} and \texttt{invariant()}. One should be careful while using it in \texttt{invariant()} or in a condition that will affect the KNX installation's state (the \textit{physical} state): the formal verification would fail if ANY possible value of the \texttt{app\_state} leads to an invalid state after running \texttt{iteration()} even if this case should not occur because of previous calls to \texttt{iteration()} that would have set the values.

\newpage

\hypertarget{app-example}{%
\subsubsection{App example}\label{app-example}}

In \texttt{main.py}:

\begin{minted}{python}
from instances import app_state, BINARY_SENSOR, SWITCH


def invariant() -> bool:
    # The switch should be on when the binary sensor is on 
    # or when INT_0 == 42, off otherwise
    return ((BINARY_SENSOR.is_on() or app_state.INT_0 == 42) and \ 
    SWITCH.is_on()) or (not BINARY_SENSOR.is_on() and \ 
    not SWITCH.is_on())


def iteration():
    if BINARY_SENSOR.is_on() or app_state.INT_0 == 42:
        SWITCH.on()
    else:
        SWITCH.off()
\end{minted}

This application sets a switch "on" (a lamp for example) when a binary
sensor is in ``on'' state (a push-button for example) or when the
\texttt{INT\_0} value equals 42. The \texttt{invariant()} ensures that
if these conditions are met, the switch is always in "on" state and if
they are not, it is set to "off".

\hypertarget{running-apps}{%
\subsection{Running apps}\label{running-apps}}

To run all the installed apps (with runtime verification enabled, see Section \ref{runtime-verification}):

\begin{enumerate}
\def\labelenumi{\arabic{enumi}.}
\tightlist
\item
  In ETS\footnote{\url{https://www.knx.org/knx-en/for-professionals/software/ets-professional/}}, import the file \texttt{assignments/assignment.csv} to create the
  group addresses, then assign to each communication object the right
  group address as presented in \texttt{assignments/assignment.txt}. The name of the group address should help to understand to which device it is meant to be assigned.
\item
  In ETS, do a basic configuration of the devices to make them have the
  correct basic behaviour (the amount of configuration depends on the
  particular device)
\item
  Execute \texttt{svshi\ run\ -a\ address:port}, where \texttt{address} is the
  KNX IP gateway address and \texttt{port} is the KNX IP gateway port.
\end{enumerate}

SVSHI logs which apps have been called during execution and which
telegrams have been received. You can find the logs in \texttt{logs/}.

For more details on how apps are run, see
\protect\hyperlink{execution}{Section \ref{execution}}.

\hypertarget{what-kind-of-productivity-benefits-could-one-expect}{%
\section{What kind of productivity benefits could one
expect}\label{what-kind-of-productivity-benefits-could-one-expect}}

With SVSHI, one can use any Python library to \textbf{develop any kind of app}: machine learning and external services can be leveraged in a \textbf{few lines} of Python. This kind of system complexity is either not achievable or very time-consuming in an ETS-only setting. Updates are \textbf{easier}, \textbf{faster} and \textbf{safer} to do, as the user just needs to modify the Python script, and the modifications are \textbf{formally verified}. In a larger perspective, one can expect \textbf{fewer errors} during the configuration of a KNX installation and \textbf{less time spent on ETS}, since SVSHI does the heavy lifting for the user.

Moreover, the SVSHI system provides a runtime verifier module. This module verifies during the execution that the invariants cannot be violated. If an app does a modification that violates at least one invariant, the modification is not applied to the KNX installation and the app (or all apps, depending on the error) is killed. As this verification works on real values known at runtime, it can detect problems with \texttt{unchecked} functions too, even if the functions' results violate post-conditions the developer added. This gives a superior safety level and can catch \textbf{all} errors.

\protect\hyperlink{how-do-we-solve-them}{Section \ref{how-do-we-solve-them}} contains more
details on this subject.


\hypertarget{whats-new-in-svshi}{%
\chapter{What's new in SVSHI?}\label{whats-new-in-svshi}}

This chapter discusses the \textbf{novelties} introduced by SVSHI.

SVSHI is one of the first platforms that allow running Python applications in smart infrastructures. It provides the highest level of abstraction available at the moment for programming KNX systems, and it extends KNX possibilities beyond what was previously achievable: any behavior and complex integration are now possible.

Applications are decoupled from the physical system they are installed on and can be developed independently. Only at installation time, the binding between devices used in the app and real physical devices is established.

Additionally, while preserving performance and developer productivity, SVSHI formally verifies applications, ensuring their invariants are always preserved. App developers need no verification expertise, and the verification process does not require their assistance.

Furthermore, formal verification has been used before for smart homes, but it was never applied to Python apps running on KNX. Moreover, symbolic execution was seldom used with Python. SVSHI is then one of the rare applications that leverage it.

\hypertarget{how-does-svshi-work}{%
\chapter{How does SVSHI work?}\label{how-does-svshi-work}}

In this chapter, we first describe how SVSHI is structured, concluding
with an in-depth explanation of the platform's internal implementation.

\hypertarget{structure}{%
\section{Structure}\label{structure}}

\begin{figure}[H]
\begin{center}\includegraphics[scale=0.32]{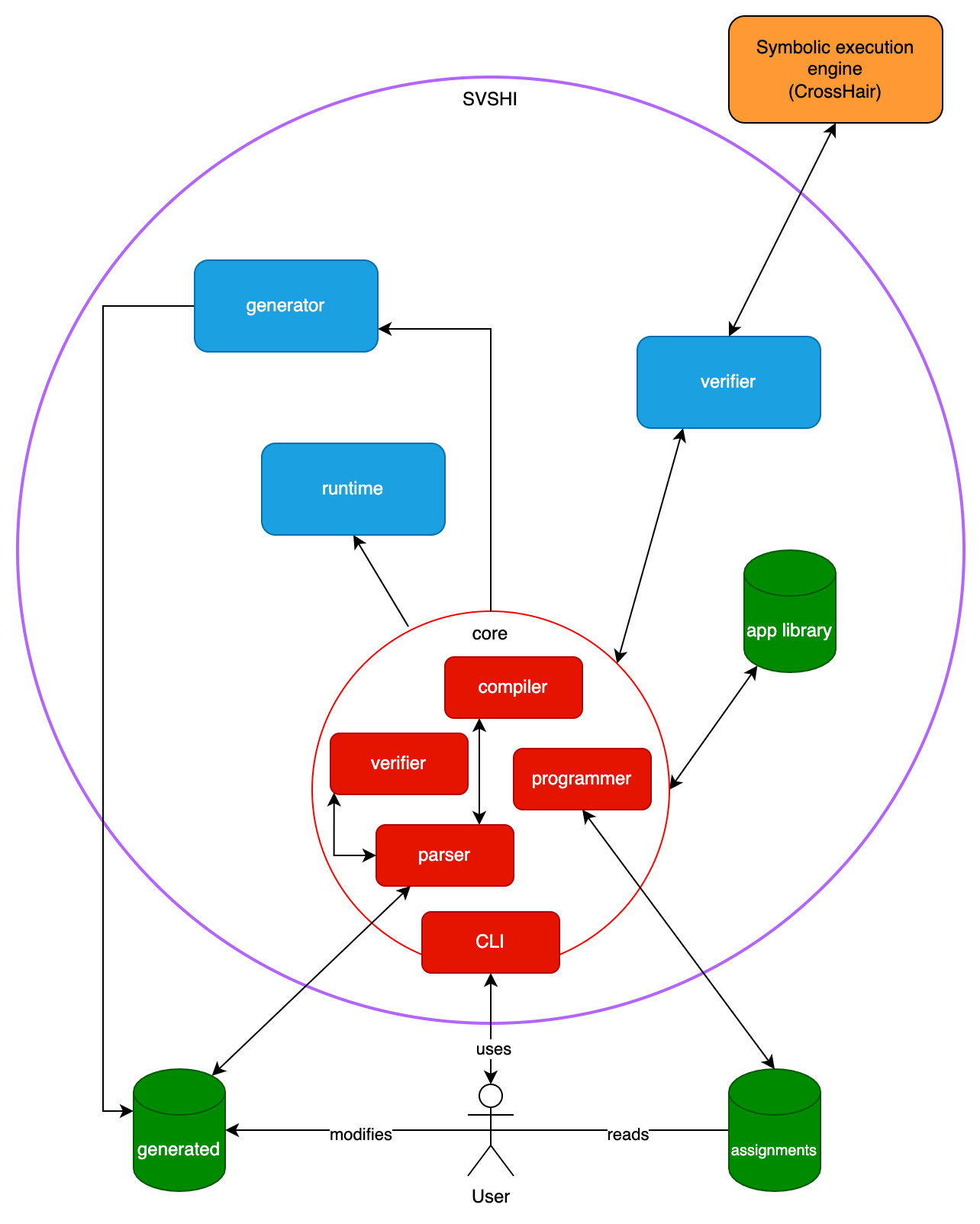}
\caption{SVSHI's structure.}
\label{structure-fig}
\end{center}
\end{figure}

SVSHI is made of different \textbf{Python} and \textbf{Scala} modules.
In Figure \ref{structure-fig}, the Python modules are in blue and the Scala ones in red. Datastores are denoted in green.

We chose Python for the \texttt{generator}, \texttt{verification} and
\texttt{runtime} modules since we need to perform code generation, AST
manipulation, code execution and symbolic execution on Python files.
Scala is used for all other tasks.

The user interacts with the infrastructure through a
\textbf{command-line interface} (CLI) which is the only entry-point. She
also has access to the \texttt{generated} folder where apps that have
been generated and in the process of being installed are stored together
with their bindings, and to the \texttt{assignments} folder where group
addresses assignments are saved for later use on ETS.

Installed apps, compiled and verified, are stored in the private
\texttt{app\_library} and can be modified or deleted only using the CLI.

\hypertarget{app-installation-pipeline}{%
\section{App installation
pipeline}\label{app-installation-pipeline}}

To install an application on SVSHI, the following tasks are performed:

\begin{itemize}
\tightlist
\item
  First, the Compiler generates a configuration of the physical devices for the applications to work (and some other information for the Verifier and KNX programmer)
\item
  Then, the Verifier performs formal verification on the applications and the devices' configuration
\item
  If the verification is successful, the KNX programmer passes this configuration to the KNX system. At this point, the new application is installed
\item
  If the verification does not pass, the current applications are kept as is and the developer must review the new application(s) before trying again
\end{itemize}

\hypertarget{implementation}{%
\section{Implementation}\label{implementation}}

In this section we describe the implementation of the different
components of the system.

\hypertarget{app-generation}{%
\subsection{App generation}\label{app-generation}}

The \textbf{app generator} produces a Python app skeleton given an app
name and the app prototypical structure JSON file.

Before executing the app generator, the user needs to create the file
containing the list of the devices the app uses, as explained later
\protect\hyperlink{4211-Prototypical-structure}{in Section \ref{prototypical-structure}}.

To execute the generator, we run
\texttt{svshi\ generateApp\ -d\ devices.json\ -n\ app\_name}, where the
first argument is the \emph{absolute} path to the prototypical structure
file and the second one (\texttt{app\_name} in the example) is the name
of the app to create. The name has to follow the same rules as for
Python modules: short, all-lowercase names. Underscores can be used if
it improves readability.

The generator first parses the JSON file, extracting the list of devices
(and failing if it has the wrong format). Then, it generates the app in
several steps:

\begin{enumerate}
\def\labelenumi{\arabic{enumi}.}
\tightlist
\item
  It creates a new directory in \texttt{generated}, if it does not
  already exist with the given name of the app.
\item
  It generates the file \texttt{instances.py} containing the devices'
  instances with their names capitalized and the app state instance.
\item
  It generates the file \texttt{\_\_init\_\_.py} in each sub-directory
  of the generated app, to make sure every folder is a Python module and
  can be used as such in imports. This step is a safety precaution in
  case the user has already started writing the app before generating
  it, which is not recommended but allowed.
\item
  It copies the predefined app skeleton to the newly generated app. The
  skeleton comprises the devices' models (\texttt{models} sub-module),
  the \texttt{requirements.txt} file, and the \texttt{main.py} file.
\item
  It generates the \texttt{multiton.py} file, placing it in the
  \texttt{models} folder. This file contains a decorator that makes the
  decorated class a multiton: only a limited number of instances can
  exist. If a device instance with a given name already exists, it is
  returned instead of creating a new object. All devices' classes are
  multitons.
\item
  It \textbf{moves} the given prototypical structure file to the newly
  generated app, renaming it \texttt{app\_prototypical\_structure.json}.
\item
  Finally, it adds import statements in \texttt{main.py} for the newly
  created instances and the app state instance, so that the user can
  directly use them in \texttt{invariant()} and \texttt{iteration()}.
\end{enumerate}

Part of the skeleton code is present to let the user use linters and run
the code locally if needed. The code is then replaced during the
installation process and is therefore not used by the SVSHI runtime
component (see Section \ref{code-manipulation}).

\hypertarget{prototypical-structure}{%
\subsubsection{Prototypical
structure}\label{prototypical-structure}}

This JSON file is written by the programmer that wants to develop an application. It represents the prototypical devices that the app needs along with their types. It also specifies whether the app is
\emph{privileged} or not
(\texttt{"permissionLevel":\ "privileged"\ \textbar{}\ "notPrivileged"}).
A privileged app overrides the behavior of the non-privileged ones at
runtime.

Moreover, the \texttt{timer} attribute can be used to run the
application even though the physical state has not changed. The app thus
becomes a \textbf{periodic} app.

\begin{itemize}
\tightlist
\item
  If \texttt{timer\ ==\ 0} the application runs only when the physical
  state of the devices it uses changes.
\item
  If \texttt{timer\ \textgreater{}\ 0} the application runs when the
  physical state changes AND every \texttt{timer} seconds.
\end{itemize}

The \texttt{files} attributes is used to indicate files that the app
needs to work properly. These files must be at the root of the
application project (next to \texttt{main.py}).

Once the app is generated, it is moved to the generated apps' folder.

Here is an example of a prototypical structure file:

\begin{minted}{json}
{
  "permissionLevel": "notPrivileged",
  "timer": 60,
  "files": ["file1.txt", "file2.png"],
  "devices": [
    {
      "name": "name_of_the_instances",
      "deviceType": "type_of_the_devices"
    }
  ]
}
\end{minted}

The \texttt{name} is used as the instance name in the Python app that is
generated. It should then be unique in a given app, and should follow
the Python variables naming conventions: no whitespaces nor numbers. The
\texttt{deviceType} should be supported by SVSHI.\\
At the moment, the supported devices are:

\begin{itemize}
\tightlist
\item
  \textbf{Binary sensors} (deviceType = ``binary'')
\item
  \textbf{Temperature sensors} (deviceType = ``temperature'')
\item
  \textbf{Humidity sensors} (deviceType = ``humidity'')
\item
  \textbf{CO2 sensors} (deviceType = ``co2'')
\item
  \textbf{Switches} (deviceType = ``switch'')
\end{itemize}

\hypertarget{compilation}{%
\subsection{Compilation}\label{compilation}}

The compilation is the part of the process of installing an application onto the system that produces the ETS configuration for the devices. Conceptually, it reads the applications and the bindings between physical and prototypical devices and outputs a configuration for the KNX devices and some files that are used by the verifier and the KNX programmer.

For now, the compilation that SVSHI performs happens in two steps. During the first one, the compiler generates a new file containing the bindings between physical communication objects and prototypical ones and the file containing the parsed physical structure. During the second phase, the compiler reads the binding file filled by the developer and assigns group addresses to each physical device communication object used at least by one application.

Let us describe more in-depth these two phases.

During the first phase, the compiler generates a file containing bindings between physical and prototypical devices. The bindings file is a JSON file that contains one object for each application that is installed or that is being installed. This object in turn contains one object for each prototypical device used by the application. Each of these devices objects contains the name and the type of the device and, more importantly, one integer for each communication object it exposes. This integer is an ID that is used to map the communication object to a physical one (each physical communication object has some ID in the \texttt{physical\_structure.json} file the compiler also produces). The developer must fill this file with the corresponding IDs before continuing. \\
One important aspect is the compiler's behaviour when generating this bindings file in case some applications are already installed. Indeed, it would not make sense to erase all existing bindings when installing a new application. However, if the devices installed in the physical installation have changed, it does not make sense to keep potentially stale mappings. \\
Therefore, the compiler checks whether the passed physical structure at compile time (i.e., the ETS project file) produces a physical installation identical to the one stored in the currently installed application library or not. If the physical structures are identical, currently existing bindings are kept and one or more new objects for the application(s) being installed are added. The developer has only to fill the bindings for the applications being installed. If the physical structures are different, all bindings are generated from scratch, old IDs are replaced by \texttt{-1} and the developer has to fill everything again, even for previously installed applications.

During the second phase, the compiler uses this bindings file that the developer filled to assign group addresses to physical devices' communication objects. Doing so, these group addresses are also assigned to prototypical devices communication objects that are mapped to physical ones. \\
The compiler assigns a group address only to communication objects that are indeed mapped to at least one prototypical device.
The compiler then generates a file containing the group addresses, the corresponding type in Python, and the corresponding KNX datatype, which will be used by the runtime module. \\
The compiler also returns values that are used by the verifier to continue the process.

\hypertarget{verification}{%
\subsection{Verification}\label{verification}}

Verification is part of SVSHI's DNA. We want to verify as much as
possible that the applications and the bindings are correct.

\subsubsection{Background}

Here we expose some background needed to understand the verification done on the applications.

\hypertarget{symbolic-execution}{%
\paragraph{Symbolic execution}\label{symbolic-execution}}

Let us start with some background about symbolic execution.

Symbolic execution is a different way of executing a program. Most interpreters run programs with concrete values (an integer would have the value 42, for example). When symbolically executing a piece of software, each variable takes a symbolic value which is a range of values that this variable could take in a concrete execution. This way, a symbolic execution explores multiple (sometimes even all) possible concrete execution paths at once.

To explore paths, a symbolic execution engine adds at each branch a 
\emph{path constraint} that represents the conditions that must hold for the symbolic values of variables for that path to be followed. An execution path in the program is feasible if the set of all paths constraint (called the \emph{path condition}) is satisfiable.

\begin{figure}[H]
\begin{center}\includegraphics[scale=0.65]{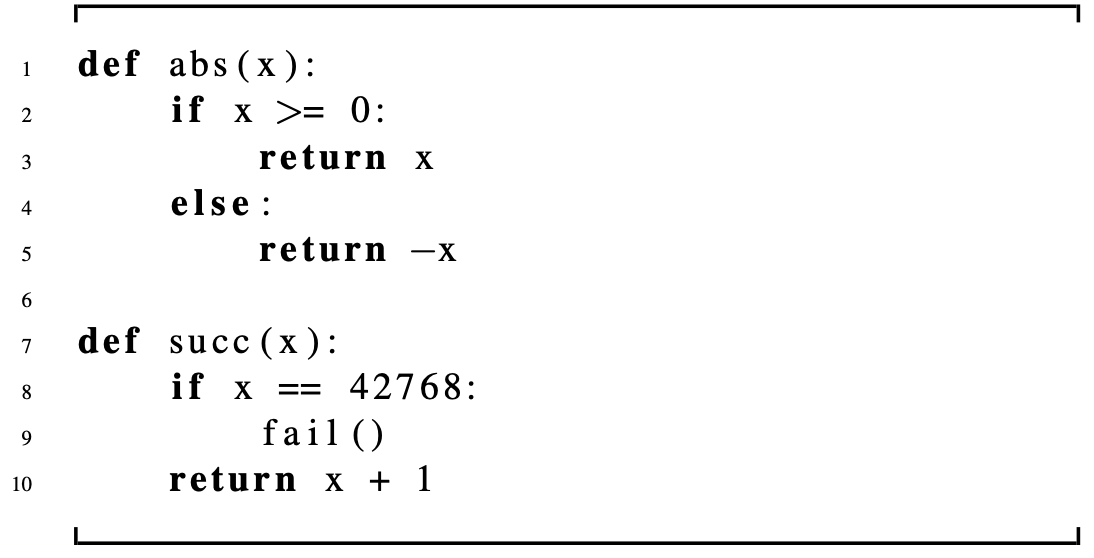}
\caption{Absolute value and faulty successor functions \cite{bruni2011}.}
\label{symbex-example}
\end{center}
\end{figure}

Let us take an example. In Figure \ref{symbex-example}, the \texttt{abs} function has one
branch. The symbolic execution will generate two path constraints, one
for the \texttt{then} branch which is \texttt{x\ \textgreater{}=\ 0}, and
one for the \texttt{else} branch which is \texttt{x\ \textless{}\ 0}.
This program also has two paths and both have a path condition that is
satisfiable.

A symbolic execution engine will choose different satisfiable paths
until the whole paths space is covered.

By using symbolic execution, it is possible to find faulty paths even if
the probability of taking them during concrete execution is extremely
low.

For example, again in Figure \ref{symbex-example}, in the \texttt{succ} function, the
symbolic execution will generate the path condition
\texttt{x\ ==\ 42768} and immediately detects that the path has a
satisfiable path condition and therefore will directly find this faulty
path. Whereas a concrete execution with a random value for argument
\texttt{x} only has 1 chance over $2^{32}$ of finding this faulty path.

\hypertarget{crosshair}{%
\paragraph{CrossHair}\label{crosshair}}

CrossHair is a Python library that allows
to analyse a Python program \cite{crosshair}. It is based on the idea developed by Bruni et Al. in their paper \cite{bruni2011}. They propose a way of implementing symbolic execution engines with few lines of code by using the actual runtime engine of the language. In a nutshell, the main idea is to leverage primitive operation dispatching to implement proxy values that will be passed around in place of the concrete values (e.g., \texttt{int}, \texttt{bool}, \ldots). Python implements everything as methods' calls: for example, \texttt{a\ +\ b} becomes
\texttt{a.\_\_add\_\_(b)} so it is possible to ``hijack'' this method
dispatching to replace primitive types with proxies. Doing so, it becomes possible to execute a function passing these proxy values instead of concrete values as arguments and record how these values are used.

Then, the engine can construct path constraints and path conditions for the program and use an SMT solver to explore paths and find counterexamples.

CrossHair proposes different ``mode'' of executions: \texttt{check},
\texttt{cover}, \texttt{diffbehavior} and \texttt{watch}. \\
\texttt{watch} is an interactive version of \texttt{check} that runs in a terminal while one is coding in her favorite editor and gives live feedback and information every time she saves the file.

We present each mode, starting with \texttt{diffbehavior} and
\texttt{cover} that we do not use for SVSHI and then \texttt{cover} that
is indeed useful for us.

In \texttt{diffbehavior} mode, given two functions, CrossHair tries to
find a set of arguments that make the two functions return different
values. In other words, it tries to find whether two functions are
identical in behaviour or not.

Using \texttt{cover} mode, CrossHair uses symbolic execution and the SMT
solver to find inputs that lead to high op-code-level test coverage.

In \texttt{check} mode, CrossHair uses symbolic execution to find
whether \emph{contracts} hold or not for functions. Contracts are annotations in Python docstrings and are pre- and post-conditions expressed as booleans expressions. The pre-conditions can restrict the range of values of the arguments and the post-conditions express what should hold at the end of the function execution (the special value \texttt{\_\_return\_\_} can be used to represent the value returned by the function). CrossHair then tries to find counterexamples, i.e. arguments' values that satisfy the pre-conditions but violate the post-conditions. If no such counterexamples are found after a timeout or if all \emph{discovered
paths} are covered, CrossHair considers the contracts to be valid.
Crosshair outputs a special message if contracts are \emph{valid over
all paths} meaning that all discovered paths were analysed and not that
the timeout was hit.

One caveat of CrossHair is the way it discovers paths. Indeed, as
CrossHair discovers paths only by \textbf{executing} the function with these
special proxy values as arguments, it only discovers feasible paths
for programs that are \emph{deterministic}. That means that branches
should not depend on external values other than arguments to be
discoverable. Also, if a program has side effects, it could cause
problems as they would be visible during the verification process.
Crosshair does not perform any AST analysis or similar, it just
executes the program with special proxy values as arguments: it performs \textit{concolic} execution.

CrossHair also requires the code to be typed. This is not a
limitation in our case as most of the code is generated so we add types
automatically. We only require the developers to type \texttt{unchecked}
functions manually.

CrossHair supports variables of type \texttt{int}, \texttt{bool},
\texttt{str} and \texttt{dict} at least \cite{crosshair-how}. We observe that it also supports
\texttt{float}. CrossHair also supports \texttt{dataclasses} and custom
classes if the attributes have one of the aforementioned types. During
our experiments, we observe that counterexamples concerning
\texttt{dict} were not found in almost all cases. We think that the
universe of values is too large, accounting for the fact that Python
dictionaries are heterogeneous and can have keys and values of any type
(even different ones in the same dictionary).

\hypertarget{crosshair-in-svshi-and-code-modification}{%
\subsubsection{CrossHair in SVSHI and code
modification}\label{crosshair-in-svshi-and-code-modification}}

We develop here what we formally verify about applications written for SVSHI and how we modify them for the verification process to work.

First of all, we need to define what are the properties that should be
verified about the applications. As described in the
\protect\hyperlink{421-app-generation}{section about app generation},
the developer writes mainly 2 functions in an application: \texttt{iteration()}
and \texttt{invariant()}. The \texttt{iteration()} function is the
function that modifies the state of the installation (i.e., what
implements the behaviour) while the \texttt{invariant()} is a function
that returns a \texttt{bool} and represents an invariant about the
physical installation that must stay valid at all times.

When performing the verification, we have 2 sets of applications: the already installed one(s) and the one(s) being installed. The main idea is that we verify that the execution of the \texttt{iteration()} function of any application on a \textit{valid state} (i.e., the installation is in a state in which all \texttt{invariant()} functions of all applications of both sets are verified) returns a possibly different but still \textit{valid} state. This condition is more conservative than it could be. Indeed, a particular set of applications may be rejected even though the execution could be valid. This would occur if the execution of the \texttt{iteration()} functions in a particular order leads to a valid state but the state is in an \textit{invalid state} between two functions. We decide that this case should be rejected as well because of the event-based nature of our system. Indeed, it would lead to invalid states if the execution order changes. It also helps during the verification process using CrossHair and we will detail how. It is therefore possible to verify \texttt{iteration()} functions independently from each other.

An important thing to note at this point is that we cannot verify that, given an \textit{invalid} state, an app produces a valid state. Indeed, for an app to produce a valid state from an invalid one, the invalid one must be from a particular subset of the universe of all states. This represents the \textit{functionality} of the app.

\newpage

For example, let us take an application that turns a light on when a presence detector detects someone. The app would look like this:

\begin{minted}{python}
def invariant():
  if PRESENCE.is_on():
    return LIGHT.is_on()
  else:
    return not LIGHT.is_on()

def iteration():
  if PRESENCE.is_on():
    LIGHT.on()
  else:
    LIGHT.off()
\end{minted}

Here, at some point in the execution, the presence will be \texttt{on} because someone has just arrived but the light will be \texttt{off} because it is exactly \textit{this} app that should set it to \texttt{on}. Thus, the state is invalid before app execution but valid after. The described scenario however cannot be verified using CrossHair because only some of the invalid states lead to valid ones after executing the apps. Some of the invalid states will stay invalid and this is normal behaviour. Therefore, to catch these errors, we implemented the runtime verification (Section \ref{runtime-verification}).

\texttt{iteration()} and \texttt{invariant()} functions are using instances of classes that represent the devices of the installation and that are communicating with KNX through the network. Given the nature of CrossHair, the code cannot use values that have to be read over the network. Moreover, as explained in the section about CrossHair (Section \ref{crosshair}), the code must not have side effects. Therefore, we need to modify the code written by the user to make it verifiable. We list and explain in detail all the modifications that we operate on the code.

First of all, we need a way to pass the state of the installation and the app state as arguments. Indeed, we want to define contracts on the states and we want the states to be symbolically represented to explore all cases. We define a class \texttt{PhysicalState} that holds the state of the installation as abstract values. For example, if the physical installation contains a push button, a light, and a temperature sensor, the \texttt{PhysicalState} will hold 2 \texttt{bool} values (representing the current state of the light and the push button) and 1 \texttt{float} value (representing the temperature measured by the temperature sensor). This \texttt{PhysicalState} is defined as a \texttt{dataclass} in Python. We do the same for the \texttt{AppState} which is a \texttt{dataclass} with a finite number of values for each supported basic type. We modify the Python code of the \texttt{iteration()} and \texttt{invariant()} functions using the Python AST module\footnote{\url{https://docs.python.org/3/library/ast.html}} to take an instance of \texttt{PhysicalState} and an instance of \texttt{AppState} as arguments. We also modify the code to pass the \texttt{PhysicalState} instance to all functions that act on devices, so that modifying the state of a device changes the instance of \texttt{PhysicalState} and getting the state of a device returns the value held by it.

We then modify \texttt{iteration()} to return a \texttt{dict} that contains the modified instances of \texttt{PhysicalState} and \texttt{AppState} so that contracts can be written about them. We return a \texttt{dict} with, as keys, the strings that contain the respective names of arguments of the \texttt{invariant()} functions. By doing so, we can call \texttt{invariant()} functions more easily with the \texttt{fun(**dict\_with\_named\_args)} notation\footnote{\url{https://book.pythontips.com/en/latest/args_and_kwargs.html\#usage-of-kwargs}}.

As explained in the previous section about CrossHair (Section \ref{crosshair}), functions passed to it must be \textit{pure}, i.e. have \textit{deterministic} behaviour and no side effects. However, we want applications to be able to use external services. We, therefore, decided to ask developers to write all the code that contains external calls in functions whose names are prefixed with \texttt{unchecked}. These functions must have an explicit return type for the verification to work. During the code modification step, we pass to the \texttt{iteration()} function one new argument for each of those functions, of the corresponding type (if the return type is not \texttt{None}, in which case the call is just replaced by \texttt{None} directly). In this way, the values returned by those functions are represented by symbolic values during the verification and all cases are explored. To aid verification and avoid false negatives, developers can add \texttt{contracts} in the form of post-conditions to these \texttt{unchecked} functions. Developers must however be very careful with these contracts because, if they are not respected at runtime by the external call, errors might occur even though the formal verification passed. \texttt{invariant()} functions cannot use \texttt{unchecked} functions.

Now that the \texttt{iteration()} function acts on an abstract representation of the states that can be symbolically represented and on \texttt{unchecked} functions corresponding arguments, we need to add contracts that CrossHair will verify. First, we add as pre-conditions all the invariants of all applications of both sets (to represent that the passed states are valid) and the post-conditions of the \texttt{unchecked} functions. Then, we add as post-conditions all the invariants of all applications of both sets but called on the return value of the \texttt{iteration()} function, to verify that, given a valid state and unchecked functions' return values, the return state is still valid.

We then call CrossHair to verify contracts for all \texttt{iteration()}
functions. If CrossHair finds a counterexample for at least one of them,
we reject the whole set of applications being installed.

Here is an example of the \texttt{iteration()} function of an
application before (as written by the developer) and after manipulation
(as passed to CrossHair):

Before:
\begin{minted}{python}
def iteration():
    if not PRESENCE_DETECTOR.is_on() and not DOOR_LOCK_SENSOR.is_on():
        if app_state.INT_0 > 5:
            unchecked_send_message("The door at office INN319 
            is still opened but nobody is there!")
        else:
            app_state.INT_0 += 1
    else:
        app_state.INT_0 = 0
\end{minted}

After:

\begin{minted}{python}
def door_lock_iteration(app_state: AppState, physical_state: PhysicalState):
    """
    pre: door_lock_invariant(app_state, physical_state)
    pre: plants_invariant(app_state, physical_state)
    pre: ventilation_invariant(app_state, physical_state)
    post: door_lock_invariant(**__return__)
    post: plants_invariant(**__return__)
    post: ventilation_invariant(**__return__)
    """
    if not DOOR_LOCK_PRESENCE_DETECTOR.is_on(physical_state) 
     and not DOOR_LOCK_DOOR_LOCK_SENSOR.is_on(physical_state):
        if app_state.INT_0 > 5:
            None
        else:
            app_state.INT_0 += 1
    else:
        app_state.INT_0 = 0
    return {'app_state': app_state, 'physical_state': physical_state}
\end{minted}

As we can see, there are 3 applications installed in the system:
\texttt{door\_lock}, \texttt{plants} and \texttt{ventilation}, which are
the 3 prototypes detailed in
\protect\hyperlink{lab-prototypes}{Section \ref{lab-prototypes}}.

\hypertarget{appstate-tradeoff}{%
\subsubsection{AppState tradeoff}\label{appstate-tradeoff}}

As explained in \protect\hyperlink{writing-apps}{the section about
how to write apps (Section \ref{writing-apps})}, applications have access to an \texttt{AppState}
instance. This lets the applications store state that persists between
calls to \texttt{iteration()} and \texttt{invariant()}.

Given our observation about how CrossHair deals with arbitrary dictionaries, we decided to not allow an arbitrary key-value store as we originally thought. Indeed, having an arbitrary dictionary as \texttt{AppState} would have led to ineffective verification and thus bugs that would have not been caught. We thus decided to propose a register-like data structure, with a finite number of values of each type for the following types: \texttt{int}, \texttt{float}, \texttt{str}, and \texttt{bool}. The number of values can be increased in future versions of the platform. The fact that the types are known and that keys (here attribute names) are fixed helps during verification.

We think that the tradeoff is acceptable, as SVSHI gets effective verification of the applications at the cost of lesser free variable names and types.

\hypertarget{dpt-and-types}{%
\subsubsection{DPT and types}\label{dpt-and-types}}

This part of the verification concerns mainly the bindings. Developers
indeed have to fill the \texttt{apps\_bindings.json} to map physical
device communication objects to prototypical device communication
objects. For the system to work properly, these bindings must be sound. Also, we cannot formally verify their correctness.\\
Therefore, we verify the most of the compatibility we can with the
information we have.

We verify the following properties:

\begin{itemize}
\item
  \textbf{\emph{IOTypes}}\\
  \texttt{IOType} can be \texttt{in}, \texttt{out} or \texttt{in/out}.
  \texttt{in} means that the prototypical (physical respectively)
  communication object can receive values from the bus and react
  accordingly. \texttt{out} means the opposite, i.e., that the
  communication object can write values to the bus. Lastly,
  \texttt{in/out} means that the communication object can do both
  simultaneously.

  Every prototypical device defined as a \texttt{SupportedDevice}
  provides the IO type for each of its communication objects (through
  the class \texttt{SupportedDeviceBinding}). The IO type of each
  physical device's communication objects is read by the parser in the
  ETS project file. As the IO type is not always provided for physical
  devices, it can be \texttt{Unknown}.

  Compatibility is defined as follows:

\begin{center}
    \begin{tabular}{| c | c  c  c |}
    \hline
       \textbf{ Physical/Prototypical} & \textbf{In} & \textbf{Out} & \textbf{In-Out} \\
        \hline
        \textbf{In} & Yes & No & No \\
        \textbf{Out} & Yes & Yes & Yes \\
        \textbf{In-Out} & Yes & Yes & Yes \\
        \textbf{Unknown} & Warning & Warning & Warning \\
        \hline
    \end{tabular}
\end{center}

  As we abstract the physical state and run applications on the
  abstraction, it means that a prototypical device in an application can
  read a state that has only \texttt{out} type in the physical world
  (because its value is stored in the mirrored state kept by SVSHI).
  This is why \texttt{in} prototypical \textless-\textgreater{}
  \texttt{out} physical is permitted.

  With the \texttt{Unknown} type for physical devices, we cannot do more
  than giving a warning to the developer who has to be sure that
  the connection is valid.
\item
  \textbf{Python types}

  As we abstract the state of the physical installation in a mirrored state, we assign a Python type (e.g., \texttt{int}, \texttt{float}, \texttt{bool}, ...) to each group address. For this to be valid, all prototypical communication objects linked to that group address must use values of the same type.

 Each device, just as for IO Types, has the Python type of the value it would read/write on the KNX bus encoded in the corresponding \texttt{SupportedDeviceBinding} class.

 This stage then checks that all communication objects connected to the same group address have the same Python type for their values.
\item
  \textbf{KNX datatypes (or DPT)}

  This stage does the same kind of verification as the IO Types one to check that the KNX datatypes of the values that the device would read/write on the bus are compatible.

 For each binding between a physical device's and a prototypical device's communication object, we check that the KNX datatype is the same. The KNX datatype of the physical device's communication object is parsed from the ETS project and the one for the prototypical's one is encoded in the corresponding \texttt{SupportedDeviceBinding} class.

 As for the IO Types, if the KNX datatype is not known for a particular physical communication object, the \texttt{Verifier} gives a warning to the developer. Otherwise, KNX datatypes must be \textbf{equal}. Here we compare the main type only, i.e., the number of bytes and type of data, not the interpretation. That means that two floats are compatible, even if one should represent a temperature and the other a CO2 level. This difference only exists for the ETS user to read data with the corresponding units.
\item
  \textbf{Mutual KNX datatypes}

  This stage performs the KNX datatype check just as the previous one but between prototypical devices' communication objects that are linked to the same physical device's communication object.
\end{itemize}

\hypertarget{runtime-verification}{%
\subsubsection{Runtime verification}\label{runtime-verification}}

The runtime verification takes place in the \texttt{runtime} module. It
leverages the \texttt{conditions.py} file, generated by the
\texttt{verification} module and copied over to \texttt{runtime} by the
\texttt{core} module. This file contains a function,
\texttt{check\_conditions()}, that given a \texttt{PhysicalState} and
multiple \texttt{AppState} (one per installed app) instances, returns a boolean
representing whether the apps' conditions are preserved. It does so by
constructing a conjunctive boolean expression with all the apps'
\texttt{invariant()} functions.

Before calling an app \texttt{iteration()} function, SVSHI checks whether the current physical state (along with its app state) is valid, namely whether it satisfies the invariants or not. If it is valid, the app is executed and if the produced state is invalid, then the app is stopped (it will not run anymore) and the state is not updated. If instead the state before running the app is not valid, the app is executed and the check is performed later. \\
The idea here is that some states are invalid because an event just occurred in the physical world (e.g., a button was pressed or a presence was detected) and a particular app implements the functionality to make it valid again (e.g., turn a light on or open a window). Thus, other apps that are running in the same iteration as this one will receive an invalid state and produce an invalid one because they will not perform the changes that this particular app has to do to make the state valid again. This is why we can only kill apps that transform a valid state into an invalid one, and, if a state remains invalid, we have to stop all apps as we cannot know which one did not behave correctly.

Once all apps have been called, the produced states are merged and SVSHI verifies that they satisfy the invariants. If it is not the case, SVSHI
restores the physical state to the \emph{last valid physical state} and
stops all the apps. The \emph{last valid physical state} is a copy of
the physical state that SVSHI maintains internally: every time the merge
operation after executing the apps produces a valid physical state, it
stores it as the \emph{last valid physical state}.\\
When the physical state is valid, the physical updates are propagated to
KNX and the app state is updated (see
\protect\hyperlink{execution}{Section \ref{execution}}).

\hypertarget{knx-programming}{%
\subsection{KNX programming}\label{knx-programming}}

This module passes the physical devices' configuration generated by the Compiler to the KNX system through ETS. For now, the configuration only consists of group addresses assignments.

This module, at the moment, only produces a list of group addresses that can be imported directly in ETS, and a text file describing in a human-readable way the assignments. The group addresses have names that make them easy to assign to the communication objects they belong to.

The rest of the ETS configuration has to be done by hand for now. However, this module will be enhanced in the future to handle more of the work automatically.

\hypertarget{execution}{%
\subsection{Execution}\label{execution}}

SVSHI's runtime is \textbf{reactive} and \textbf{event-based}.
Applications \emph{listen} for changes to the group addresses of the
devices they use and are run on a state change (an \emph{event}). The
state transition can be triggered externally by the KNX system or by
another app, which then proceeds to notify all the other listeners.
Notable exception are apps that run every X seconds based on a timer. These apps not only react to state changes but are also executed \textit{periodically}.

\emph{Running an application} concretely means that its
\texttt{iteration()} function is executed on the current physical state
of the system and the current app state.

Apps are always run in alphabetical order in their group
(\texttt{privileged} or \texttt{notPrivileged}). The non-privileged apps
run first, then the privileged ones: in such a way privileged
applications can override the behavior of non-privileged ones. 

This order of execution is abstract, as apps could be theoretically run in parallel; however, the merging of the new copies of the physical state they produce must then be carefully done to preserve the order.

\newpage

This execution model has been chosen for its \textbf{ease of use}: users
do not need to write \texttt{while} loops or deal with synchronization
explicitly.

\hypertarget{code-manipulation}{%
\subsubsection{Code manipulation}\label{code-manipulation}}

The app skeleton provided to the user when she generates an app contains classes modeling the devices the app can use. However, these classes are just stubs given for code auto-completion and linting, since they are not the ones executed by SVSHI. The \texttt{verification}
module does not simply modify the code to ease verification, as
explained in
\protect\hyperlink{crosshair-in-svshi-and-code-modification}{Section
\ref{crosshair-in-svshi-and-code-modification}}: it also generates the file \texttt{runtime\_file.py} that is
used during execution by the \texttt{runtime} module (see next section, Section \ref{runtime}).
This file is similar to \texttt{verification\_file.py}, as it contains:

\begin{itemize}
\tightlist
\item
  The states: both \texttt{AppState} and \texttt{PhysicalState}
\item
  The actual device classes: one per each device instance in an app, as
  devices of the same type might access different group addresses (in a
  single app or across different apps)
\item
  The actual device instances used
\item
  The \texttt{iteration} functions of each installed app
\end{itemize}

However, compared to the file generated for verification, it does not need to contain contracts or invariants, as they are not used during execution.
Moreover, \texttt{unchecked} functions that perform side effects and all user-defined imports are kept. The file is internal to SVSHI and is not meant to be read by the developer.

Here is an example of the \texttt{iteration()} function of the
application \texttt{app\_one} before (as written by the developer) and
after AST manipulation (as passed to \texttt{runtime}):

Before:

\begin{minted}{python}
from instances import app_state, BINARY_SENSOR, SWITCH

def iteration():
    if BINARY_SENSOR.is_on() or app_state.INT_0 == 42:
        unchecked_send_email('test@test.com')
        SWITCH.on()
    else:
        SWITCH.off()
\end{minted}

\newpage

After:

\begin{minted}{python}
APP_ONE_BINARY_SENSOR = Binary_sensor_app_one_binary_sensor()
APP_ONE_SWITCH = Switch_app_one_switch()

def app_one_iteration(app_state: AppState, physical_state: PhysicalState):
    if APP_ONE_BINARY_SENSOR.is_on(physical_state
        ) or app_state.INT_0 == 42:
        app_one_unchecked_send_email('test@test.com')
        APP_ONE_SWITCH.on(physical_state)
    else:
        APP_ONE_SWITCH.off(physical_state)
    return {'app_state': app_state, 'physical_state': physical_state}
\end{minted}

Here is an example of device class:

\begin{minted}{python}
@dataclasses.dataclass
class PhysicalState:
 GA_0_0_1: bool
 GA_0_0_2: bool

class Switch_app_one_switch():
    def on(self, physical_state: PhysicalState):
        physical_state.GA_0_0_2 = True

    def off(self, physical_state: PhysicalState):
        physical_state.GA_0_0_2 = False

    def is_on(self, physical_state: PhysicalState) -> bool:
        return physical_state.GA_0_0_2
\end{minted}

As we can see, the classes are modified to update the physical state instead of directly sending and reading data from KNX. This feature is leveraged at runtime, which is the subject of the next section.

\hypertarget{runtime}{%
\subsubsection{Runtime}\label{runtime}}

In more detail, the \texttt{runtime} module first initializes the group addresses listeners, i.e. the apps in \texttt{app\_library}, installing their requirements and saving their \texttt{iteration} functions (coming from the \texttt{runtime\_file.py} that has been generated by the \texttt{verification} module and copied over by \texttt{core}, as previously explained). Then, from each app, it parses \texttt{addresses.json}, reading the XKNX\footnote{\url{https://xknx.io}} DPT classes associated with each group address: they are used to decode and encode the data coming from the bus. Then, the physical state is initialized by connecting to KNX and reading all group addresses: it represents a local copy of the KNX system state. Finally, the module starts listening for KNX telegrams. Whenever a telegram is received for a given address, the local physical state is updated and the listeners are notified.

For what concerns app states, they are initialized with default values:
\texttt{0} for \texttt{int}, \texttt{0.0} for \texttt{float},
\texttt{False} for \texttt{bool} and \texttt{""} for \texttt{str}.

Note that, since the state (both physical and app-specific) can be updated concurrently by periodic apps and by apps reacting to incoming telegrams, a lock is used to \textbf{synchronize} the modifications and avoid deadlocks and re-orderings.

As explained previously, all apps are \textbf{executed} in the \textbf{same way} and in the \textbf{same order}: first non-privileged apps in alphabetical order, then privileged ones again in alphabetical order.
The execution happens in the following manner: first, we generate copies of the physical state and the app state for each app that needs to be executed. Then, sequentially, each app is run \textit{on its own copy of the physical and app state}; if invariants are not preserved and the physical state was valid before the execution, we prevent the app from running again (see Section \ref{runtime-verification}), otherwise, the app state is updated and the physical state is saved to later propagate it. Once the new physical states resulting from all the apps' executions have been recorded, we \textbf{merge} them together with the old state before execution, and we update the local physical state. Finally, we write to the KNX bus for just the final values given to the updated fields of the state, and we notify the listeners of the changes.

This execution plan has the advantage of sending \textbf{less data} through the KNX bus, as apps are executed in batches and state updates are coalesced to only take into account the latest value, discarding all intermediate states. It also avoids sending unnecessary telegrams that would ask a device to change its state to its current value (e.g., send an "on" command to a light that is already in the "on" state). \\
Moreover, apart from the initialization phase, \textbf{no reads} are sent to the KNX bus. \\
Furthermore, another benefit of keeping a local copy of the physical state is that apps can read the state of devices that are usually \textbf{write-only} in a KNX-only usage, such as some actuators.

On the other hand, the distributed nature of KNX is lost, as most logic is \textbf{centralized} in SVSHI.

\hypertarget{prototype-evaluation}{%
\chapter{Prototype \& evaluation}\label{prototype-evaluation}}

In this section, we first present the prototypes we developed with SVSHI. Then, we evaluate what has been accomplished together with its limitations and we conclude with a discussion about future work and improvements.

\hypertarget{lab-prototypes}{%
\section{Lab prototypes}\label{lab-prototypes}}

To demonstrate SVSHI's usage, we developed \textbf{3 app prototypes} in
our lab, using a small KNX installation:

\begin{itemize}
\tightlist
\item
  An app to track \textbf{soil moisture level} of plants that sends
  messages on the lab \textbf{Slack} workspace if the level is below the
  recommended threshold.
\item
  An app that monitors the \textbf{lock state} of our office's door and
  that sends messages on \textbf{Slack} according to the
  \textbf{presence} detector installed in the room as well: the alert is
  sent if there is no one in the room and the door is not locked.
\item
  An app that monitors the conference room \textbf{presence}, the
  \textbf{CO2 level} and the lab \textbf{Google Calendar} and that turns
  the \textbf{ventilation} "on" or "off". Ventilation is turned "on":

  \begin{itemize}
  \tightlist
  \item
    when someone is in the room
  \item
    when a meeting is scheduled in the next 15 minutes
  \item
    during a meeting (with respect to the calendar)
  \item
    when the CO2 level is above 900 ppm
  \end{itemize}
\end{itemize}

The \textbf{KNX system} is made of various devices connected to the bus:

\begin{itemize}
\tightlist
\item
  A KNX power supply.
\item
  A KNX USB interface.
\item
  A KNXnet/IP interface.
\item
  A switch.
\item
  A binary sensor.
\item
  A humidity sensor with a plant probe.
\item
  A door lock sensor (that connects to the binary sensor to communicate over KNX).
\item
  A room presence detector which can also measure the CO2, the
  temperature and the brightness levels.
\end{itemize}

\begin{figure}[H]
\begin{center}\includegraphics[scale=0.25]{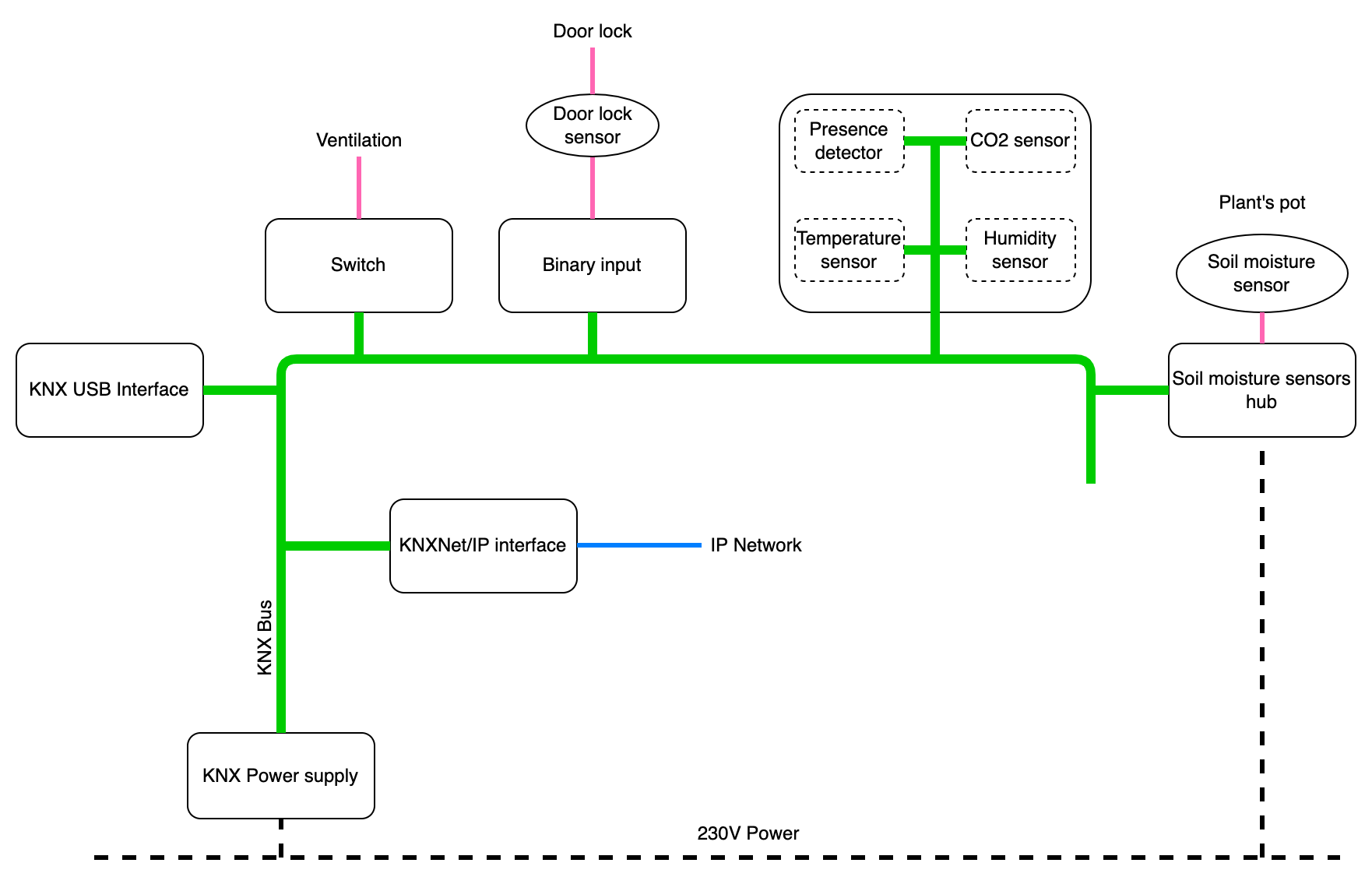}
\caption{KNX prototype installation.}
\label{prototypes-diagram-fig}
\end{center}
\end{figure}

Figure \ref{prototypes-diagram-fig} shows a technical diagram of the KNX setup we installed.

\begin{figure}[H]
\begin{center}\includegraphics[scale=0.09]{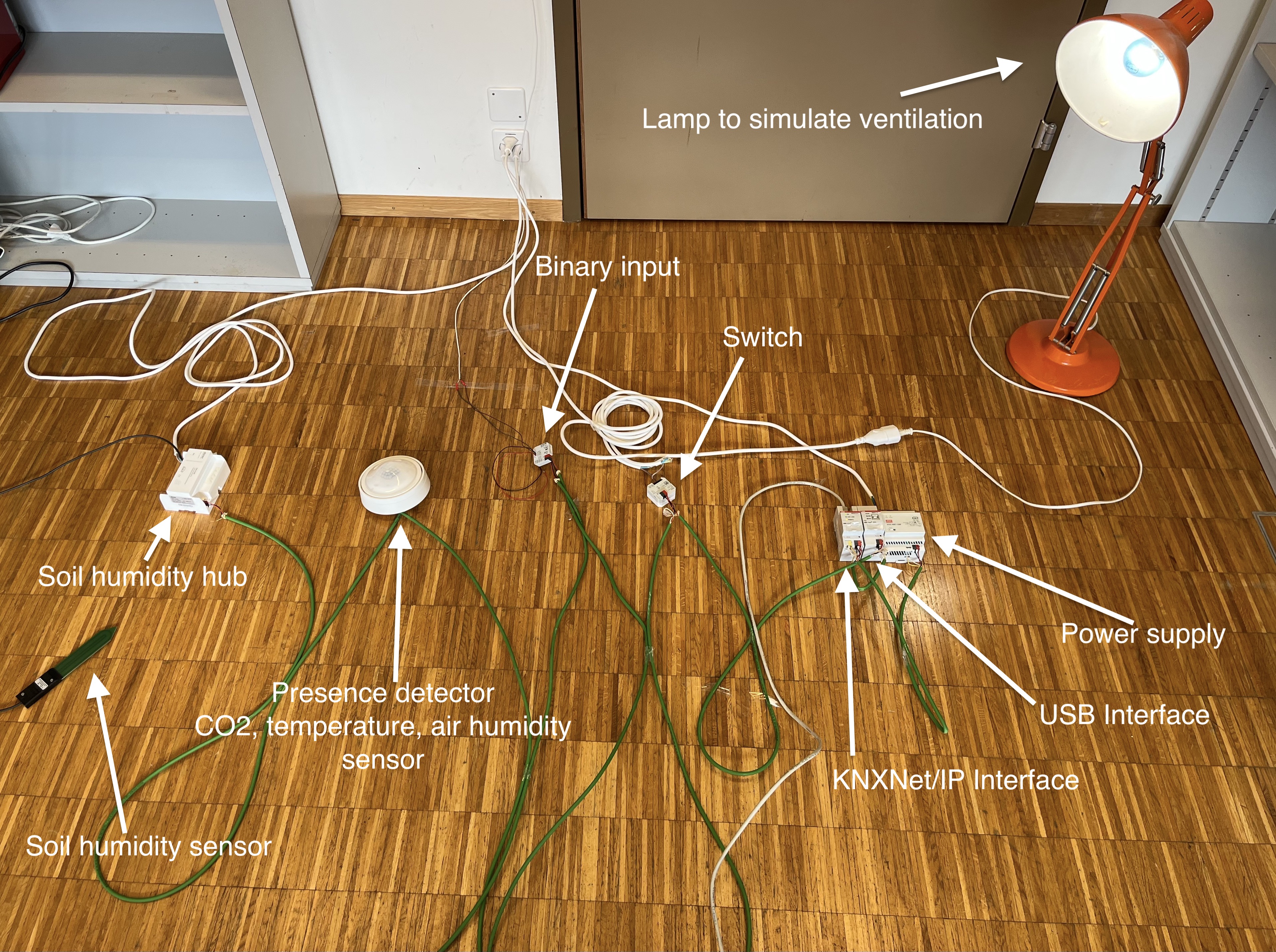}
\caption{KNX prototype physical installation.}
\label{physical-installation-photo-fig}
\end{center}
\end{figure}

\begin{figure}[H]
\begin{center}\includegraphics[scale=0.35]{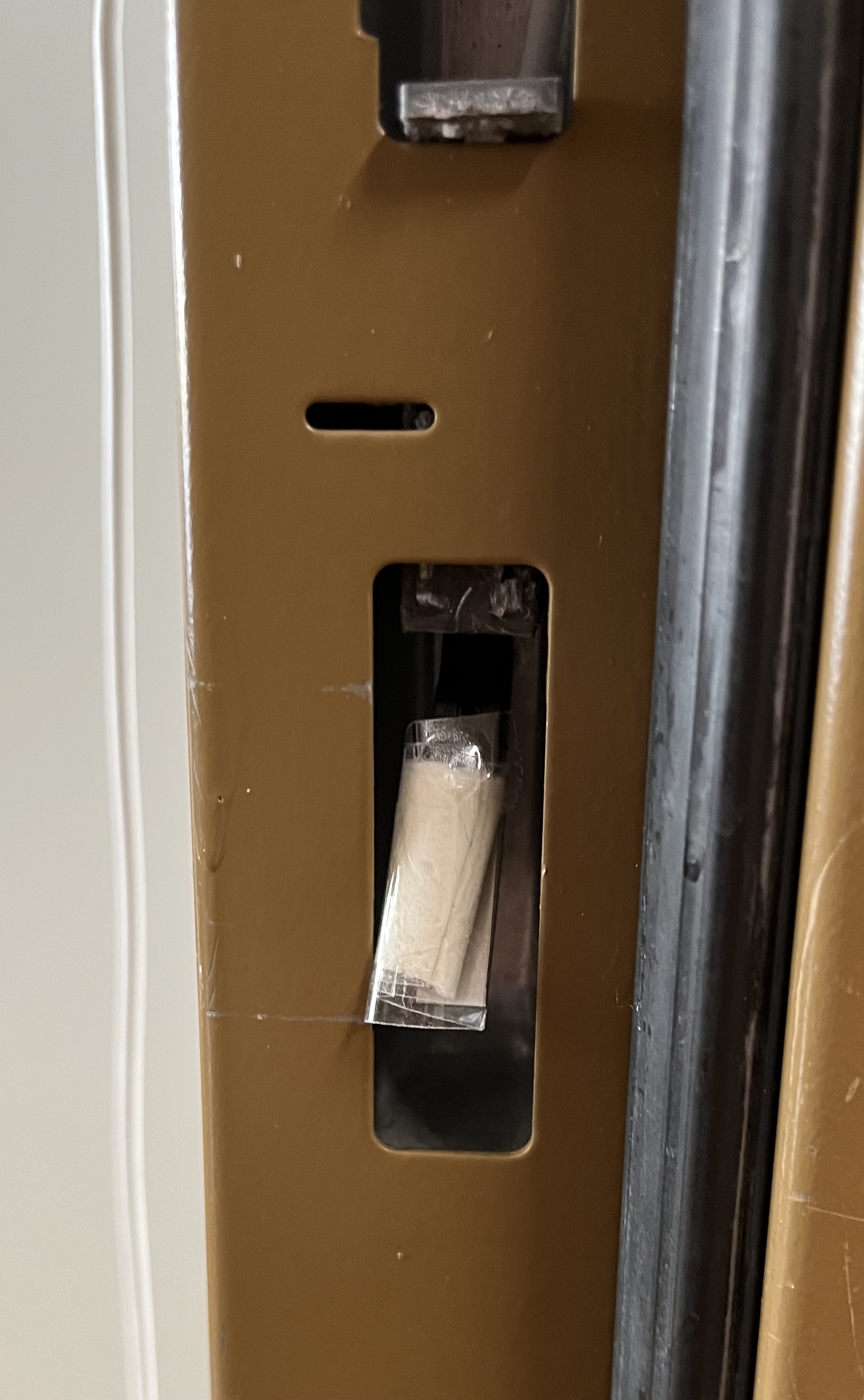}
\caption{Door lock sensor.}
\label{door-lock-photo-fig}
\end{center}
\end{figure}

In Figure \ref{physical-installation-photo-fig}, we can see a photo of the installation. Figure \ref{door-lock-photo-fig} shows the door lock sensor.

\hypertarget{accomplishments}{%
\section{Accomplishments}\label{accomplishments}}

The prototypes exemplify SVSHI's main features and accomplishments:

\begin{itemize}
\tightlist
\item
  The programming of KNX devices and the implementation of complex behavior in a few lines of Python.
\item
  The usage of external services (Google Calendar and Slack) in apps, which is not achievable with KNX alone.
\item
  The execution of multiple apps in the same system.
\item
  The automatic formal verification of the apps: the user does not need
  a background in formal verification, she just needs to write the
  invariants the apps must satisfy.
\item
  The possibility for apps to read the state of devices that are usually
  write-only in a KNX-only usage, such as actuators.
\end{itemize}

\hypertarget{limitations}{%
\section{Limitations}\label{limitations}}

SVSHI is great, but it has some important \textbf{limitations}:

\begin{itemize}
\tightlist
\item
  We lose the decentralized aspect of KNX, as most logic is
  \emph{centralized}: if the host running SVSHI goes down, the KNX
  system does not work anymore.
\item
  We do not use the full potential of the devices: in Python, we implement sometimes behaviors that could be done by the device itself on its own with the right ETS configuration. This is since two push-buttons do not necessarily have the same settings available, so SVSHI uses the most basic ones that are always available (see Section \ref{knx-devices}).
\item
  We still need to use ETS to program devices the first time an app is installed and this can be a cause of errors. However, this programming is mostly limited to group addresses assignment, which is aided by the \texttt{assignment.txt} and \texttt{assignment.csv} files that SVSHI outputs on a successful compilation.
\item
  SVSHI itself is not verified and thus stays for now in the Trusted Computing Base (TCB).
\end{itemize}

\newpage

\hypertarget{future-work}{%
\section{Future work}\label{future-work}}

In this section, potential future work and directions are discussed.

First of all, some of the limitations described previously could be overcome:

\begin{itemize}
\tightlist
\item
  SVSHI could program the devices without the need for the user to use ETS manually. During our research we came to the understanding that it is indeed possible but rather laborious; however, it could prove very useful for end-users, as they could work mostly without ETS. 2 directions are possible: either a simpler approach in which SVSHI generates XML project files to be opened in ETS directly (the user just needs to click download to program the devices), or providing alongside SVSHI an ETS application that programs the devices.
\item
  SVSHI could decentralize part of the programming to the devices.
\item
  SVSHI itself could be verified. This would, however, most likely need first a re-writing of a large part of the current platform, which was not written to be formally verified as-is.
\end{itemize}

The centralization trend could also be countered by \textbf{clustering}:
having SVSHI running on multiple hosts at once with the same state, in a
similar fashion as what \href{https://kubernetes.io/}{Kubernetes}\footnote{\url{https://kubernetes.io/}} does
with
\href{https://kubernetes.io/docs/concepts/workloads/controllers/replicaset/}{replica
Pods}\footnote{\url{https://kubernetes.io/docs/concepts/workloads/controllers/replicaset/}}.

Moreover, a relatively simple improvement would be adding
\textbf{support for more device types} than the five currently
available, such as control units, brightness sensors, etc.

Furthermore, an \textbf{app marketplace} needs to be implemented to
stimulate SVSHI's adoption and help users less keen on programming. It
could be represented, for example, as a server on the cloud, with formal
verification performed on-the-fly for app uploads and downloads
performed via the CLI.

Another important feature to provide is a \textbf{graphical user
interface} (GUI) for SVSHI to replace the current CLI, as the user still needs to update manually some JSON configuration files: this could be done graphically. SVSHI already provides an interface in \texttt{Svshi.scala} with a singleton to be used, should one want to implement the GUI in Scala. Otherwise, we suggest developing it in another programming language and calling directly the CLI. The next step in this direction would be transitioning to a system in which SVSHI is used as a web app, with no installation required.

Additionally, a \textbf{simulator} for KNX, based on real devices XML files, would be very useful for testing and development purposes. It could also be used by KNX professionals to perform demonstrations to customers, for example by building a virtual installation and running some apps. For now, the KNX Association offers a simulator they call \textit{KNX Virtual} (see Section \ref{knx-simulator}). This is a Windows program that offers some devices that can be programmed using ETS; the user can play with the GUI to change sensors states and see the reaction of the other devices. The main issue is that these devices only exist in KNX Virtual. This means that the settings are simpler (from our experience) than real devices'. Moreover, if you want to simulate a known infrastructure (e.g., your own), you cannot. For this reason, we propose to write a simulator core that takes XML files provided by manufacturers (the ones ETS is using) to simulate real devices in the local LAN. In the beginning, not all settings have to be supported, even SVSHI does not. The simulator could evolve alongside SVSHI and support new settings when it does. It could also be completely separated. A separated simulator could serve many more purposes than just being a SVSHI companion.

For what concerns the applications' execution, a potential improvement could be running apps of the same permission level \textbf{in parallel} instead of sequentially, with a finer-grained and smarter ordering than the one currently implemented. This could improve the performance of the platform, which however should not be an immediate concern with a relatively small number of apps installed. Nevertheless, it could prove beneficial once users start to include machine learning models running locally in their applications and other long-lasting operations.
In addition, it could be interesting for apps to communicate between them, either via message-passing or via a shared state. One could thus imagine hierarchies of apps (based on permission level or functionality), with high-ranked apps distributing work and coordinating "worker" apps.

Finally, there is still some work to do for what concerns apps' formal verification too: CrossHair could be modified to perform \textbf{exhaustive symbolic execution} on all paths (and not just the ones discovered by executing the code), so that SVSHI can guarantee 100\% verified apps. A completely new symbolic execution engine for Python could also be developed from scratch.

\hypertarget{why-you-should-contribute-to-svshi}{%
\chapter{Why you should contribute to
SVSHI}\label{why-you-should-contribute-to-svshi}}

SVSHI is an \textbf{open-source} project that welcomes and encourages external contributions. It represents a good opportunity to contribute to a \textbf{cutting-edge research} project, working in rapidly evolving and exciting fields such as \textbf{smart infrastructures} and \textbf{formal verification}. The platform is useful and already usable, but there is still \textbf{room for improvements}, as explained in the previous section.

\chapter{Conclusion}

In this thesis, we presented our work on formal verification in smart infrastructures, focusing on verifying through symbolic execution applications running in a KNX infrastructure with the long-term goal of providing 100\% secure and reliable smart buildings. We developed an open-source platform, SVSHI, to write, compile, verify, and run Python applications on KNX, and we exploited it to implement 3 real-life prototypes in our laboratory. The results are encouraging and show that advanced yet verified applications can be easily and swiftly developed for KNX infrastructures.


\cleardoublepage
\phantomsection
\addcontentsline{toc}{chapter}{Bibliography}
\printbibliography

%
%

\end{document}